\documentclass[apj]{emulateapj}

\shorttitle{AKARI IRC 2.5-5 $\mu$m spectroscopy of infrared galaxies}
\shortauthors{Ichikawa et al.}

\usepackage{lscape}
\usepackage{amsmath}
\usepackage{color}
\usepackage{url}
\usepackage{ulem}
\usepackage{multirow}
\usepackage{longtable}

\def\lir{$L_{\rm IR}$} 
\def\lireq{L_{\rm IR}} 
\def\lsuneq{L_{\odot}} 

\begin{document}

\title{\textit{AKARI} IRC 2.5--5~$\mu$\lowercase{m} spectroscopy of infrared galaxies over a
wide luminosity range}


\author{Kohei Ichikawa\altaffilmark{1},
 Masatoshi Imanishi\altaffilmark{2},
 Yoshihiro Ueda\altaffilmark{1},
 Takao Nakagawa\altaffilmark{3},
 Mai Shirahata\altaffilmark{3, 4},
 Hidehiro Kaneda\altaffilmark{5},
 Shinki Oyabu\altaffilmark{5}
 }
\affil{\altaffilmark{1}
Department of Astronomy, Graduate School of Science, Kyoto University,
Kitashirakawa-Oiwake cho, Kyoto 606-8502, Japan
}
\affil{\altaffilmark{2}
Subaru Telescope, 650 North A'ohoku Place, Hilo, HI 96720, USA
}
\affil{\altaffilmark{3}
Institute of Space and Astronautical Science (ISAS), Japan Aerospace Exploration Agency, 3-1-1 Yoshinodai, Chuo-ku, Sagamihara, Kanagawa 252-5210, Japan
}
\affil{\altaffilmark{4}
National Institutes of Natural Science, National Astronomical Observatory of Japan (NAOJ)
   2-21-1 Osawa, Mitaka, Tokyo 181-8588, Japan
}
\affil{\altaffilmark{5}
Graduate School of Science, Nagoya University, Furo-cho, Chikusa-ku, Nagoya, Aichi 464-8602, Japan
}
\email{ichikawa@kusastro.kyoto-u.ac.jp}

\begin{abstract}

We present the result of a systematic infrared 2.5--5~$\mu$m
spectroscopic study of 22 nearby infrared galaxies over a wide infrared
luminosity range ($10^{10} L_{\odot} < L_{\rm IR} < 10^{13}
L_{\odot}$) obtained from \textit{AKARI} Infrared Camera (IRC). The
unique band of the \textit{AKARI} IRC spectroscopy enables us to
access both the 3.3~$\mu$m polycyclic aromatic hydrocarbon (PAH) emission
feature from star forming activity and the continuum of torus-dust
emission heated by an active galactic nucleus (AGN). Applying our AGN
diagnostics to the \textit{AKARI} spectra, we discover 14 buried AGNs. The
large fraction of buried AGNs suggests that AGN activity behind
the dust is almost ubiquitous in ultra-/luminous infrared galaxies
(U/LIRGs). We also find that both the fraction and energy contribution
of buried AGNs increase with infrared luminosity from $10^{10}
L_{\odot}$ to $10^{13} L_{\odot}$, including normal infrared galaxies
with $L_{\rm IR} < 10^{11} L_{\odot}$. The energy contribution from
AGNs in the total infrared luminosity is only $\sim$ 7\% in LIRGs and
$\sim$~20\% in ULIRGs, suggesting that the majority of the infrared
luminosity originates from starburst activity. Using the PAH emission,
we investigate the luminosity relation between star formation and
AGN. 
We find that these infrared galaxies exhibit  higher star formation rates 
than optically selected Seyfert galaxies with the same AGN luminosities,
implying that infrared galaxies could be an early evolutionary phase of AGN.

\end{abstract}

\keywords{galaxies: active --- galaxies: nuclei --- infrared: galaxies}

%
\section{INTRODUCTION}

Dust emission gives us crucial information to understand both the
history of cosmic star formation in galaxies and that of supermassive
black hole (SMBH) growth in galactic centers. Intense star formation
produces a great amount of dust, which makes its activity invisible in
the ultraviolet band but visible in the infrared band. Similarly, the
central engines of active galactic nuclei (AGNs) are surrounded by
dusty ``tori'' \citep{kro86}. 
Since optical and ultraviolet lights are very easily obscured by the
torus, the most complete search for AGNs can be made by detecting hard
X-rays from the central engine or infrared emission from the heated
dust.
Utilizing X-ray data of the Chandra Deep Field South, \cite{bri12}
reported that the fraction of ``Compton-thick'' AGNs (those with
absorption column densities of $N_{\rm H} > 10^{24}$cm$^{-2}$)
increases with redshift from $z=0$ to $z>1$. This fact implies that
the level of dust-obscuration of AGN may be correlated to the star
formation activity, which has a peak at $z\sim2$ \citep{hop06b}. Indeed,
\cite{gou12} gathered all available {\it Spitzer} spectra of local
Compton-thick AGNs and found strong starburst (SB) features in their
average infrared spectrum, suggesting that SB activity and highly
obscured tori are somehow coupled each other. \cite{ich12} also
reported the same trend for buried AGNs with small opening angles of
tori \citep[``new type'' AGNs;][]{ued07, win09, egu09, egu11} detected
in the \textit{Swift}/BAT survey. Although hard X-rays are useful to search for
obscured AGNs, extremely heavily buried AGNs with $N_{\rm H} >
10^{24.5}$cm$^{-2}$ are difficult to be detected due to flux
attenuation by repeated Comptonization even at energies above 10 keV
\citep[e.g.,][]{ike09,bri11}.


Candidates of galaxies that host such buried AGNs are infrared
galaxies, defined as those having large infrared luminosities (\lir)
with $\lireq \ge 10^{10} \lsuneq$. Among infrared galaxies, those
 in the range $10^{11} \lsuneq \le \lireq < 10^{12} \lsuneq$ are
called luminous infrared galaxies \citep[LIRGs;][]{san96} and more
luminous ones in the range $10^{12} \lsuneq \le \lireq < 10^{13}
\lsuneq$ are called ultraluminous infrared galaxies
\citep[ULIRGs;][]{san88}. Their bolometric luminosities are dominated
by the infrared emission, suggesting that very luminous heating
sources are surrounded by dust and then the heated dust re-emits in
the infrared band. The hidden energy sources are believed to be either
SB, AGN, or both. Disentangling the energy sources of infrared
galaxies is crucial to unveil the history of dust obscured star
formation and SMBH growth in the universe \citep{lef05,got10, mag11, mur11}.

\cite{vei95, vei99, yua10} studied AGN activity in U/LIRGs on the
basis of optical observations. They showed that the fraction of AGNs
increases with infrared luminosities in the range of $10^{10} \lsuneq
\le \lireq < 10^{13} \lsuneq$. One caveat of their optical diagnostics
is that one cannot identify buried AGNs embedded in dust-tori with very
small opening angles where no significant narrow line regions can be formed.
Unlike the AGNs which are optically-classified type-2 Seyfert, such buried AGNs are
elusive to conventional optical spectroscopy \citep{mai03}.  After the
launch of \textit{Spitzer}, the AGN activity in local U/LIRGs has been
also addressed in the infrared band. \cite{nar08, nar09, nar10}
observed ULIRGs with \textit{Spitzer}/IRS 5--8~$\mu$m spectroscopy,
and investigated both the SB and AGN contributions by fitting SB/AGN
templates to the spectra. They found the increasing AGN significance
with infrared luminosity, confirming the same trend reported from the
optical studies. \cite{alo12} observed the 5--38~$\mu$m spectra of LIRGs 
with \textit{Spitzer}. They decomposed the spectra into SB and AGN
components using a SB galaxy template and clumpy torus models
\citep{nen08a, nen08b, ase09}, which are supported by observations
\citep[e.g.,][]{gan09, ich12}. They confirmed that 
the AGN energetic contribution to the total power
increases with infrared luminosity  from $10^{11} \lsuneq \le \lireq \le 10^{12}\lsuneq$.

Infrared 2.5--5 $\mu$m spectroscopy is another powerful tool to study
optically-elusive buried AGNs in U/LIRGs. One advantage of this band
is that dust extinction is much lower than in the optical
band \citep{nis09}  and is similar to that in the 5--13 $\mu$m band
\citep{lut96}.  Another great advantage is that SB and AGN activity
can be distinguished based on the spectral features.  First, strong
polycyclic aromatic hydrocarbon (PAH) emission, a pure star formation
tracer, is located at 3.3~$\mu$m in SB galaxies, while pure AGNs
exhibit PAH-feature free spectra due to the effects that 1) the X-ray
emission destroying the PAHs, and 2) the strong continuum emission
originating from AGN hot-dust diminishes this feature \citep{moo86,
ima00}.
SB galaxies
generally show large 3.3~$\mu$m PAH equivalent widths of
${\rm EW}_{\rm 3.3PAH} \sim 100$~nm, which never go down
below 40~nm. Thus, objects with ${\rm EW_{\rm 3.3PAH}} \le 40$~nm
may be classified as galaxies that harbor buried AGNs.
In SB/AGN composite galaxies, strong PAH emission can be observed
because the AGN cannot destroy the PAH molecules located in the
outer ($r > 10$~pc) region from the central engine due to 
shielding by dust and gas. However, strong AGN continuum makes EW$_{\rm 3.3PAH}$
smaller, well below $40$~nm. Using this method, many authors
reported the strong sign of buried AGNs in ULIRGs \citep{ima06,
san08, ris10} and LIRGs \citep{ima08, ima10}. Another key 
feature in the 2.5--5~$\mu$m band is the continuum slope $\Gamma$ ($F_{\nu}
\propto \lambda^{\Gamma}$). SB galaxies have blue continuum
slopes ($\Gamma \sim 0$) in this band due to the
contribution of the stellar photospheric continuum.
Conversely, galaxies with AGNs have hot-dust emission heated by
AGNs, which produces a red continuum ($\Gamma \ge 1$). This method was
first reported by \cite{ris06} and successfully applied to ULIRGs
\citep{san08, ris10} and LIRGs \citep{ima08,ima10}.
\cite{ima10} combined the two methods (${\rm EW_{\rm 3.3PAH}} \le
40$~nm and/or $\Gamma > 1$) to find buried AGNs from
a sample of U/LIRGs showing no AGN signatures in the optical spectra.
They found that the fraction of buried AGNs increases with infrared
luminosity in the range of $10^{11} \lsuneq \le \lireq \le 10^{13}
\lsuneq$.

In this paper, we expand the studies of U/LIRGs by \cite{ima10} into a
lower infrared luminosity range, by including normal infrared galaxies
(IRGs) with $10^{10} \lsuneq \le \lireq < 10^{11} \lsuneq$, where
comprehensive studies are still missing due to the limited
sample. This is because L/IRGs have spatially more extended infrared
emission ($> 1$~kpc, corresponding to $>$several arcseconds at $z \sim
0.05$) compared with ULIRGs \citep{soi01, ima11b}, and therefore slit
spectroscopy with ground based telescopes could miss their extended
emission. Slit-less spectroscopy of Infrared Camera
\citep[IRC;][]{ona07} on board \textit{AKARI} \citep{mur07} with 1$'
\times 1'$ aperture opened a new window to probe such extended
emission of L/IRGs.   
Throughout this paper, we adopt $H_{0} =
75$~km s$^{-1}$Mpc$^{-1}$, $\Omega_{\rm M} = 0.3$, and
$\Omega_{\Lambda} = 0.7$ for consistency with the previous
publications \citep{ima08, ima10}.

%
\section{TARGETS}

Our motivation is to search for buried AGNs from optically non-Seyfert infrared galaxies and
investigate their properties over a wider range of infrared luminosity than in previous studies.  
An ideal sample can be selected from unbiased catalogs of infrared galaxies 
with various infrared luminosities
at redshifts of $z<0.45$, 
where 3.3~$\mu$m PAH is detectable in the \textit{AKARI}/IRC
2.5--5.0~$\mu$m spectra.
To this end, we first gather U/LIRGs with high far-infrared fluxes
from \textit{IRAS} catalogs.
Our ULIRGs are
mainly selected from the bright ULIRG catalog by \citet{kla01}, 
consisting of 41~sources 25 out of which are optically non-Seyfert
galaxies. These ULIRGs have \textit{IRAS} 60~$\mu$m
fluxes $\ge 3$~Jy and far-infrared (40--120~$\mu$m) luminosities
$L_{40-120~\mu {\rm m}} > 10^{12} L_{\odot}$\footnote{This
 luminosity is based on the paper of \cite{kla01}, where $H_0 = 50$~km s$^{-1}$Mpc$^{-1}$.
Therefore, in our study with $H_0 = 75$~km s$^{-1}$Mpc$^{-1}$,  some galaxies reach $L_{\rm IR} = 10^{11.5} L_{\odot}$.}.
In addition to these sources, we also gather 54 fainter, local ($z<0.45$) optically
non-Seyfert ULIRGs from the literature to increase the sample size.
The LIRGs are mainly taken from the LIRG
catalog of \cite{car88}, which is based on the \textit{IRAS} Bright Galaxy
catalog. The catalog consists of 61 sources with the criteria of
\textit{IRAS} 60~$\mu$m fluxes $\ge 5.4$~Jy and far-infrared
(40--400~$\mu$m) luminosities $L_{40-120~\mu {\rm m}} > 10^{11} L_{\odot}$. 
We select 41 optically non-Seyfert galaxies out of this sample.
To further increase the number of L/IRG targets, 
we also utilize the catalog by \citet{spi02}.
They compiled 76 \textit{ISO} detected
sources in the \textit{IRAS} 12~$\mu$m galaxy catalog,
which gathered all galaxies at Galactic latitudes of $|b| > 25^{\circ}$
down to the \textit{IRAS} 12~$\mu$m flux limit of $\ge 0.22$~Jy
\citep{rus93}.
Since AGN tori have peak emission at the MIR band, 12~$\mu$m galaxy
catalog could tend to include more AGNs than in far-infrared selected
samples. Indeed, a majority of the \citet{spi02} catalog (64 out of 76 sources)
are optically Seyfert galaxies. Therefore, we gather the 12 optically
non-Seyfert starburst galaxies from it.
In summary, we select a sufficiently large number of non-Seyfert
infrared galaxies (132 sources) as our ``parent'' sample to be 
followed-up with \textit{AKARI}. The redshifts of all the sources in
the above catalogs are distributed within $z<0.45$.

\begin{figure}[tbp]
\begin{center}
\includegraphics[angle=0,scale=0.6]{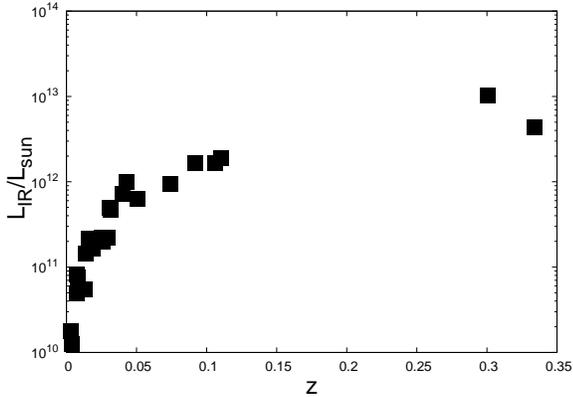}
\\
\caption{
Redshift versus infrared luminosity plot for the 22 
infrared galaxies in our sample.
\label{fig-1}}
\end{center}
\end{figure}

%
\section{OBSERVATIONS AND DATA REDUCTION}

Observation was conducted with the IRC infrared spectrograph
\citep{ona07} on board \textit{AKARI} \citep{mur07}. The spectra in
the 2.5--5~$\mu$m band were taken with the NG grism mode. This mode
achieves an effective spectral resolution of $R \sim 120$ at
3.6~$\mu$m, which is sufficient for detecting and tracing the profile
of $3.3$~$\mu$m PAH emission. The IRC has a $1 \times 1$ arcmin$^2$
window with a pixel scale of $1.46 \times 1.46$~arcsec$^2$.
All the data were taken as a part of
the \textit{AKARI} mission program called ``AGNUL'' (PI:
T.~Nakagawa). The observation settings were same as those 
described in Chapter~3 of \citet{ima10}. 
One to seven pointings
were assigned for each source according to the brightness. 
The total on-source exposure time was $\sim 6$~minutes per each pointing.
We used IRCZ4 (phase 3; post liquid-He mission) observing mode, where
one pointing is composed of eight or nine independent frames
\citep{ona07}. This mode successfully removes cosmic-ray 
contaminations even for sources observed by a single pointing. For
sources observed with multiple pointings, we combined all the data to
achieve the best signal-to-noise (S/N) ratio after excluding the data sets with
particularly bad quality. We often had to discard data
observed in later epochs of \textit{AKARI}'s phase 3 operation due to
the increasing background signal.

Spectral analysis was conducted in a standard manner by using the IDL
data reduction package for {\it AKARI} IRC spectra\footnote{The
software package is available through the \textit{AKARI} data
reduction webpage
\url{http://www.ir.isas.jaxa.jp/AKARI/Observation/}. For the details
of the data analysis, see \cite{ohy07}.}. 
The dark-subtraction, linearity correction, and flat-field correction
were performed using this IDL package. Some infrared galaxies with
low luminosities ($L_{\rm IR} < 10^{12} L_{\odot}$) have spatially
extended emission. Hence, we varied the aperture size for spectral
extraction according to the actual signal profile of each source. The
background was estimated from both sides of the spectral dispersion
direction of the target and was subtracted from the source. The
IDL package for \textit{AKARI} also performed wavelength and 
absolute flux calibration.
The accuracy of the wavelength calibration is $\sim
1$~pixel ($\sim 0.01$~$\mu$m), and that of the flux calibration 
is $\sim10$\% around the central wavelength of the spectra
and up to $\sim20$\% at the edges close to 2.5~$\mu$m and 5.0~$\mu$m.

Out of the 132 sources in our parent sample, we observed
37 objects in total from 2008 June until 2010 February with
\textit{AKARI}.
After the epoch, \textit{AKARI} fell into the stage where the data highly
 suffered from the significant background emission. As the result,
we could not observe the remaining 95 targets.
Also, due to \textit{AKARI}'s Sun-synchronous polar orbit, higher
visibilities were achieved for objects at higher ecliptic latitudes.
These \textit{AKARI}'s observational constraints make it difficult for
us to construct a uniformly flux-limited sample 
over the entire sky with
\textit{AKARI}. However, the observable targets are essentially
randomly selected from the parent sample with no
selection biases regarding the physical nature of the galaxies.
Hence, we regard that our sample has no obvious bias.

Next, we also set the criterion that the averaged S/N ratio for each
spectral channel must be higher than 5
in order to apply our spectral deconvolution analysis.
Among the 37 targets observed with \textit{AKARI}, 
five sources were discarded due to their low S/N ratio and the other 10
sources were too faint to be detected. 
14 out of 15 non-detected sources are fainter ULIRGs gathered from the literature. 
These sources have 60~$\mu$m flux~$= 1.1$~Jy, which is roughly 10 times fainter
 than that of the detected sources ($f_{60\mu{\rm m}}=11.8$~Jy).
Supposing that their infrared spectral energy distribution does not change drastically 
from that of the detected sources, we can regard that the most of the non-detected
 sources are intrinsically too faint to be observed with \textit{AKARI}/IRC.
Figure~1 shows the redshift distribution of the finally selected 22 infrared galaxies.

\begin{deluxetable*}{lr*{7}c}
\tablecaption{Basic Information of Our Sample\label{tbl1}}
\tablewidth{0pt}
\tablehead{
\colhead{Name} & \colhead{ObsID} & \colhead{$z$} & \colhead{$f_{12}$} & \colhead{$f_{25}$} & \colhead{$f_{60}$}
 & \colhead{$f_{100}$} & \colhead{$\log L_{\rm IR}$} & \colhead{Components}\\ 
\colhead{(1)} & \colhead{(2)} & \colhead{(3)} & \colhead{(4)} & \colhead{(5)} & \colhead{(6)} & \colhead{(7)}
 & \colhead{(8)} & \colhead{(9)} 
}
\startdata
ESO~286-IG19     & 1122176-001 & 0.043 &   0.28 &   1.90 &  12.71 &  10.58 & 12.00 &      2comp\\
IC~5135          & 1120236-001 & 0.016 &   0.63 &   2.14 &  16.67 &  26.27 & 11.33 &      2comp\\
IRAS~03068-5346  & 1122105-001 & 0.074 &   0.25 &   0.20 &   3.43 &   3.92 & 11.98 &      2comp\\
IRAS~03538-6432  & 1122106-001 & 0.30 &   0.25 &   0.25 &   0.96 &   1.54 & 13.01 &      2comp\\
IRAS~10494+4424  & 1122096-001 & 0.092 &   0.25 &   0.25 &   3.36 &   5.60 & 12.22 &      2comp\\
IRAS~17028+5817  & 1122101-001 & 0.11 &   0.25 &   0.08 &   2.49 &   4.05 & 12.22 &      2comp\\
IRASF~07353+2903 & 1120200-001 & 0.33 &   0.08 &   0.10 &   0.22 &   0.64 & 12.63 &      2comp\\
MCG~+02-04-025   & 1122123-001 & 0.031 &   0.37 &   1.46 &  11.13 &  10.29 & 11.67 &      2comp\\
MCG~+08-23-097   & 1122130-001 & 0.029 &   0.25 &   0.53 &   5.08 &   8.10 & 11.34 &      2comp\\
MCG~+10-19-057   & 1122132-001 & 0.031 &   0.40 &   1.92 &  11.35 &  10.81 & 11.70 &      2comp\\
Mrk~1490         & 1122134-001 & 0.026 &   0.25 &   0.86 &   6.21 &   8.38 & 11.30 &      3comp\\
Mrk~551          & 1122171-001 & 0.050 &   0.26 &   0.80 &   4.63 &   6.13 & 11.80 &      2comp\\
Mrk~848          & 1122136-001 & 0.040 &   0.32 &   1.52 &   9.38 &  10.26 & 11.86 &      2comp\\
NGC~2339         & 1120230-001 & 0.0074 &   0.53 &   2.11 &  18.96 &  32.24 & 10.69 &      3comp\\
NGC~2388         & 1120231-001 & 0.014 &   0.49 &   1.99 &  16.21 &  23.09 & 11.15 &      3comp\\
NGC~4102         & 1120232-001 & 0.0028 &   1.45 &   6.83 &  47.37 &  68.50 & 10.25 &      2comp\\
NGC~4194         & 1122091-001 & 0.0083 &   0.86 &   4.36 &  22.79 &  25.94 & 10.88 &      3comp\\
NGC~4818         & 1120234-001 & 0.0036 &   0.85 &   3.88 &  20.26 &  26.55 & 10.10 &      3comp\\
NGC~520          & 1120229-001 & 0.0076 &   0.78 &   2.83 &  31.52 &  48.40 & 10.91 &      3comp\\
NGC~6285         & 1122138-001 & 0.019 &   0.30 &   0.44 &   7.43 &  23.58 & 11.21 &      3comp\\
NGC~838          & 1122173-001 & 0.013 &   0.59 &   1.83 &   0.40 &  17.94 & 10.74 &      3comp\\
ZW~453.062       & 1122140-001 & 0.025 &   0.29 &   0.57 &   7.59 &  11.04 & 11.34 &      2comp\\
\enddata
\tablecomments{Table~1 summarize the basic information of our targets. 
(1) object name: 
(2) observation ID of \textit{AKARI}:
(3) redshift: (4)--(7) 12, 25, 60, and 100~$\mu$m {\it IRAS} flux density in the unit of Jansky (Jy):
 (8) total infrared (8--1000~$\mu$m) luminosity in units of solar luminosity ($L_{\odot}$), calculated from $L_{\rm IR} = 2.1 \times 10^{39} 
\times D({\rm Mpc})^2 \times (13.48f_{12} + 5.16f_{25} + 2.58f_{60} + f_{100})$ \citep{san96}: 
 (9) number of continuum components adopted to fit each spectrum: 
``2comp'' and ``3comp'' correspond to the model spectra given by 
equation~(6) and (7) in section~4.4, respectively. In a case of  ``3comp'', the main flux contribution originates from stellar and \textsc{Hii} components 
with little contribution from dust component. This is why for ``3comp'' sources the dust component (dashed-dot line) is hardly seen in Figure~2.
 }\\
\end{deluxetable*}

%
\section{AGN/SB SPECTRAL DECOMPOSITION and BURIED AGN DIAGNOSTICS}

For understanding the properties of buried AGNs in infrared galaxies, it
is crucial to analyze the infrared spectra by decomposing SB and AGN
components. The 2.5--5.0~$\mu$m band covered by \textit{AKARI} IRC has
unique advantages for detecting AGN signatures. As introduced in
Section~1, one promising method for finding buried AGNs is the strength
(equivalent width) of the 3.3~$\mu$m PAH emission as proposed by
\cite{ima00}. Pure SB galaxies show generally strong 3.3~$\mu$m PAH
emission with EW$_{\rm 3.3PAH} \sim 100$~nm because plenty of
SB-produced UV photons excite PAH molecules \citep{mou90}. By contrast,
if an AGN resides in the galaxy, the PAH feature becomes weaker,
EW$_{\rm 3.3PAH} < 40$~nm \citep{ima00}. \cite{ima08, ima10} also used
the red continuum slope ($\Gamma > 1$ for $F_{\nu} \propto
\lambda^{\Gamma}$) for finding buried AGNs. In this paper, we develop
the previous works not only to find buried AGNs but also to
quantitatively estimate their energy contribution by performing 
detailed spectral fitting with a continuum model consisting of a stellar
component and dust components from AGN and/or SB. Emission and
absorption line features are also included in the model.

\subsection{AGN-heated dust component}

The AGN emission in the near- to mid-infrared originates from the
radiation reprocessed by hot dust in the torus. 
Recent studies of dusty torus models reproduce
the observed nuclear mid-infrared emission well by assuming that the
torus media have a clumpy structure \citep[e.g.,
][]{nen08a,nen08b,ram09,ram11,alo11}. However, some authors mentioned
that there are difficulties in fitting their near-infrared spectral
energy distribution (SED) \citep{sta12,lir13,vid13}. This is due to
the complicated degeneracies including a possible existence of an
extra hot-dust ($\sim 1500$~K) component originating from
the vicinity of the AGN \citep{kis11}, and/or contamination from the
host galaxies, and/or the extinction of the torus emission by
interstellar matter in the host galaxies.
In this work, we take the simplest approach to approximate the torus
dust emission by a single blackbody component, considering our limited wavelength range.
Presumably, the dusty torus has a continuous dust temperature distribution and
produces emission peaked around $\sim$10~$\mu$m. 
Hence, our single blackbody fit should trace its peak temperature at $\sim$300~K.
Accordingly, we set a conservative upper limit on its temperature as $T_{\rm dust} < 800$~K \citep{oya11}.

\subsection{Stellar and Starburst Component}

The stellar photospheric emission in the host galaxy 
contributes to the total radiation
in the 2.5--5.0~$\mu$m spectral range. This component produces a
decreasing continuum flux at $\ge 1.8$~$\mu$m \citep{saw02}. 
We first determine the temperature $T^{(\rm stellar)}$ by fitting the 2MASS
$J, H,$ and $K_{\rm s}$ photometric data with a blackbody model, and
fix the temperature at its best-fit value when analyzing the
\textit{AKARI} 2.5--5.0~$\mu$m spectra.
For Mrk~551 and Mrk~848, we cannot find any 2MASS fluxes.
Therefore, we use the $J, H,$ and $K$ photometric data from the literature \citep{spi95} instead of the 2MASS data.
SB produces a large amount of dust, which generally has lower
temperatures than those of the AGN-heated dust, typically $T^{(\rm
dust)} < 100$~K \citep{san96}. 
As already suggested by \cite{saw02} and many other studies,
such cold dust 
with $T^{(\rm dust)} < 100$~K does not significantly contribute
to the 2.5--5.0~$\mu$m spectra, and therefore we ignore the contribution
from the SB dust emission.
However, some SB galaxies show excess emission from very hot dust
with a temparature of $\sim 10^3$~K, 
requring at least two blackbody components.
This very hot dust component of $\sim 10^3$~K would not originate 
from the dusty torus, from which blackbody radiation of $\sim$300~K 
is generally expected (see Section~4.1.).
One possible origin is the emission mainly heated by massive stars in
\textsc{Hii} regions \citep{hun02,lu03, sie93}. If such emission is required
from the data, we model it with a single blackbody with a temperature
of $T_{\textsc{Hii}} \sim 1000$~K in a range of $800 < T_{\textsc{Hii}} < 1200$~K
as suggested by \cite{hun02}.

\setlength{\tabcolsep}{0.03in} 
\begin{deluxetable*}{lc*{8}c}
\tabletypesize{\scriptsize}
\tablecaption{Fitting Properties and AGN Signs Obtained from the {\it AKARI} 2.5--5.0~$\mu$m
Spectroscopy\label{tbl2}}
\tablewidth{0pt}
\tablehead{
\colhead{Name} &
\colhead{EW$_{\rm 3.3PAH}$} &
\colhead{$T^{\rm (stellar)}$} &
\colhead{$T^{(\rm \textsc{Hii})}$} &
\colhead{$T^{\rm (dust)}$} &
\colhead{AGN} &
 \colhead{AGN} &
 \\
 &&&&& \colhead{sign (EW)} & \colhead{sign (red)}  \\
\colhead{(1)} & \colhead{(2)} & \colhead{(3)} & \colhead{(4)} & \colhead{(5)} & \colhead{(6)} & \colhead{(7)}
}
\startdata
ESO~286-IG19      & $  68.9 \pm  5.2$ & $2786\pm   94 $  & $    \cdots   $  & $358\pm   4 $  & N & Y \\
IC~5135           & $  47.3 \pm  1.2$ & $2929\pm   37 $  & $    \cdots   $  & $749\pm  20 $  & N & Y \\
IRAS~03068-5346   & $  90.4 \pm  5.7$ & $3148\pm  411 $  & $    \cdots   $  & $700\pm  82 $  & N & Y \\
IRAS~03538-6432   & $  31.9 \pm  4.9$ & $1475\pm   80 $  & $    \cdots   $  & $304\pm  15 $  & Y & Y \\
IRAS~10494+4424   & $  81.2 \pm  4.8$ & $2083\pm  144 $  & $    \cdots   $  & $415\pm  22 $  & N & Y \\
IRAS~17028+5817   & $  54.2 \pm  4.5$ & $2206\pm  199 $  & $    \cdots   $  & $366\pm  17 $  & N & Y \\
IRASF~07353+2903  & $   5.2 \pm  3.9$ & $1544\pm  154 $  & $    \cdots   $  & $249\pm  56 $  & Y & Y \\
MCG~+02-04-025    & $ 121.3 \pm  5.0$ & $2545\pm   86 $  & $    \cdots   $  & $349\pm  11 $  & N & Y \\
MCG~+08-23-097    & $  41.4 \pm  1.6$ & $2612\pm   74 $  & $    \cdots   $  & $649\pm  64 $  & N & Y \\
MCG~+10-19-057    & $  76.1 \pm  3.0$ & $2455\pm   68 $  & $    \cdots   $  & $708\pm  71 $  & N & Y \\
Mrk~1490          & $  59.7 \pm  1.6$ & $2568\pm   70 $  & $1128\pm   94 $  & $     < 100 $  & N & N \\
Mrk~551           & $  41.5 \pm  2.8$ & $2545\pm  141 $  & $    \cdots   $  & $499\pm  13 $  & N & Y \\
Mrk~848           & $  93.9 \pm  4.2$ & $3142\pm  227 $  & $    \cdots   $  & $715\pm  43 $  & N & Y \\
NGC~2339          & $  42.3 \pm  1.4$ & $2968\pm   41 $  & $1199\pm  275 $  & $     < 100 $  & N & N \\
NGC~2388          & $  55.3 \pm  1.1$ & $2715\pm   37 $  & $1177\pm   95 $  & $     < 100 $  & N & N \\
NGC~4102          & $  38.7 \pm  1.7$ & $2971\pm   44 $  & $    \cdots   $  & $739\pm  27 $  & Y & Y \\
NGC~4194          & $  78.2 \pm  1.8$ & $2986\pm   46 $  & $1192\pm   68 $  & $     < 100 $  & N & N \\
NGC~4818          & $  42.0 \pm  1.3$ & $3240\pm   61 $  & $1200\pm  217 $  & $     < 100 $  & N & N \\
NGC~520           & $  54.2 \pm  1.1$ & $2897\pm   55 $  & $1149\pm   75 $  & $     < 100 $  & N & N \\
NGC~6285          & $  59.9 \pm  2.1$ & $2940\pm   83 $  & $1173\pm  194 $  & $     < 100 $  & N & N \\
NGC~838           & $ 110.9 \pm  1.2$ & $2863\pm   41 $  & $1200\pm  185 $  & $     < 100 $  & N & N \\
ZW~453.062        & $  50.9 \pm  3.2$ & $2725\pm   55 $  & $    \cdots   $  & $649\pm  71 $  & N & Y \\
\enddata
\tablecomments{Table~2 summarizes the observed properties of the 22 sources obtained from the {\it AKARI} spectroscopy. 
(1) object name: 
(2) equivalent width of 3.3~$\mu$m PAH emission: 
(3), (4), (5) temperature of black body originated from stellar emission, \textsc{Hii} emission, and dust emission. See Section~4 for the detail: 
(6) AGN sign based on the PAH diagnostic. ``Y'' represents the source has buried-AGN sign (EW$_{\rm 3.3PAH} < 40$~nm), while ``N'' represents
no AGN sign (EW$_{\rm 3.3PAH} > 40$~nm): 
(7) AGN sign based on the 
hot-dust diagnostic. ``Y'' represents the source with $T^{(\rm dust)} > 200$~K, while ``N'' represents the source with $T^{(\rm dust)} < 200$~K.
}\\
\end{deluxetable*}

\subsection{Emission and Absorption lines}

The 2.5--5.0~$\mu$m spectra of infrared galaxies contain various
emission/absorption line features. The previous studies of the \textit{
AKARI} ``AGNUL'' program \citep{ima08, ima10} reported three PAH lines
at 3.3, 3.4, and 3.5~$\mu$m, hydrogen Brakett series at 4.05~$\mu$m
Br$\alpha$ and 2.63~$\mu$m Br$\beta$, and Pfund series at 4.65~$\mu$m
Pf$\beta$ and 3.74~$\mu$m Pf$\gamma$ as emission lines.  They also
reported absorption features at 3.1~$\mu$m by H$_2$O ice, at
4.29~$\mu$m by CO$_2$, and at 4.67~$\mu$m by CO. Because PAH emission
is believed to originate from solid state molecules, the line profile
becomes Lorentzian \citep{yam13}. For simplicity, however, we model
all the lines including those of PAH by Gaussian profiles, following
\cite{ima10}. 
The central wavelength and line width are fixed within a spectral resolution
 at appropriate values reported in the literature.

\begin{turnpage}
\setlength{\tabcolsep}{0.05in} 
\begin{deluxetable*}{lc*{10}c}
\tabletypesize{\scriptsize}
\tablecaption{Observed Properties Obtained from the {\it AKARI} 2.5--5.0~$\mu$m
Spectroscopy\label{tbl2}}
\tablewidth{0pt}
\tablehead{
\colhead{Name} &
  \colhead{$f_{\rm BB}^{(\rm stellar)}$} &
    \colhead{$f_{\rm BB}^{(\rm \textsc{Hii})}$} &
      \colhead{$f_{\rm BB}^{(\rm dust)}$} &
   \colhead{$f_{\rm 3.3PAH}$} &
   \colhead{$\tau_{\rm ice}$} &
    \colhead{$\log L_{\rm BB}^{(\rm stellar)}$} &
       \colhead{$\log L_{\rm BB}^{(\rm \textsc{Hii})}$} &
          \colhead{$\log L_{\rm BB}^{(\rm dust)}$} &
     \colhead{$\log L_{\rm 3.3PAH}$} &
      \colhead{$L_{\rm BB}^{(\rm dust)}/L_{\rm IR}$} \\
\colhead{(1)} & \colhead{(2)} & \colhead{(3)} & \colhead{(4)} & \colhead{(5)} & \colhead{(6)} & \colhead{(7)}
 & \colhead{(8)} & \colhead{(9)} & \colhead{(10)}  & \colhead{(11)}
}
\startdata
ESO~286-IG19      & $ 112\pm   4$ & $    \cdots $ & $ 338\pm  21$ & $  1.58\pm  0.12$ & $  0.13\pm  0.05$ & $44.62\pm0.01$ & $    \cdots  $ & $45.10\pm0.03$ & $41.77\pm0.03$ & $0.329\pm0.020$\\
IC~5135           & $ 856\pm  12$ & $    \cdots $ & $ 137\pm   9$ & $  8.66\pm  0.21$ & $       <   0.02$ & $44.64\pm0.01$ & $    \cdots  $ & $43.84\pm0.03$ & $41.65\pm0.01$ & $0.085\pm0.005$\\
IRAS~03068-5346   & $  54\pm   7$ & $    \cdots $ & $   7\pm   2$ & $  0.75\pm  0.05$ & $  0.20\pm  0.07$ & $44.80\pm0.06$ & $    \cdots  $ & $43.87\pm0.13$ & $41.95\pm0.03$ & $0.020\pm0.005$\\
IRAS~03538-6432   & $  16\pm   1$ & $    \cdots $ & $  29\pm   6$ & $  0.15\pm  0.02$ & $       <   0.09$ & $45.60\pm0.02$ & $    \cdots  $ & $45.85\pm0.09$ & $42.58\pm0.07$ & $0.180\pm0.034$\\
IRAS~10494+4424   & $  25\pm   2$ & $    \cdots $ & $  19\pm   4$ & $  0.56\pm  0.03$ & $  1.02\pm  0.08$ & $44.66\pm0.03$ & $    \cdots  $ & $44.54\pm0.09$ & $42.02\pm0.03$ & $0.055\pm0.010$\\
IRAS~17028+5817   & $  21\pm   2$ & $    \cdots $ & $  22\pm   4$ & $  0.26\pm  0.02$ & $  0.30\pm  0.12$ & $44.70\pm0.04$ & $    \cdots  $ & $44.74\pm0.09$ & $41.82\pm0.04$ & $0.085\pm0.015$\\
IRASF~07353+2903  & $  15\pm   2$ & $    \cdots $ & $  16\pm  15$ & $  0.02\pm  0.01$ & $  0.38\pm  0.21$ & $45.66\pm0.07$ & $    \cdots  $ & $45.71\pm1.30$ & $41.74\pm0.62$ & $0.308\pm0.292$\\
MCG~+02-04-025    & $ 114\pm   4$ & $    \cdots $ & $ 340\pm  52$ & $  3.20\pm  0.13$ & $  0.27\pm  0.06$ & $44.35\pm0.01$ & $    \cdots  $ & $44.82\pm0.07$ & $41.80\pm0.02$ & $0.366\pm0.055$\\
MCG~+08-23-097    & $ 153\pm   5$ & $    \cdots $ & $  19\pm   5$ & $  1.41\pm  0.06$ & $  0.39\pm  0.04$ & $44.42\pm0.01$ & $    \cdots  $ & $43.51\pm0.13$ & $41.39\pm0.02$ & $0.038\pm0.010$\\
MCG~+10-19-057    & $  54\pm   2$ & $    \cdots $ & $  10\pm   3$ & $  1.11\pm  0.04$ & $  0.08\pm  0.06$ & $44.02\pm0.01$ & $    \cdots  $ & $43.26\pm0.11$ & $41.33\pm0.02$ & $0.010\pm0.002$\\
Mrk~1490          & $  89\pm   7$ & $  33\pm   4$ & $    \cdots $ & $  2.04\pm  0.06$ & $      \cdots   $ & $44.07\pm0.04$ & $43.64\pm0.05$ & $    \cdots  $ & $41.43\pm0.01$ & $    \cdots   $\\
Mrk~551           & $ 113\pm   6$ & $    \cdots $ & $ 106\pm  10$ & $  1.39\pm  0.09$ & $       <   0.03$ & $44.77\pm0.02$ & $    \cdots  $ & $44.74\pm0.04$ & $41.86\pm0.03$ & $0.228\pm0.020$\\
Mrk~848           & $ 117\pm   9$ & $    \cdots $ & $  30\pm   5$ & $  2.43\pm  0.11$ & $  0.18\pm  0.03$ & $44.58\pm0.03$ & $    \cdots  $ & $43.99\pm0.06$ & $41.90\pm0.02$ & $0.035\pm0.005$\\
NGC~2339          & $ 526\pm  33$ & $ 101\pm  25$ & $    \cdots $ & $  5.68\pm  0.19$ & $      \cdots   $ & $43.74\pm0.03$ & $43.02\pm0.12$ & $    \cdots  $ & $40.77\pm0.01$ & $    \cdots   $\\
NGC~2388          & $ 534\pm  26$ & $ 120\pm  13$ & $    \cdots $ & $  8.88\pm  0.17$ & $      \cdots   $ & $44.30\pm0.02$ & $43.65\pm0.05$ & $    \cdots  $ & $41.52\pm0.01$ & $    \cdots   $\\
NGC~4102          & $1629\pm  26$ & $    \cdots $ & $ 235\pm  20$ & $ 13.10\pm  0.56$ & $  0.20\pm  0.03$ & $43.40\pm0.01$ & $    \cdots  $ & $42.56\pm0.04$ & $40.30\pm0.02$ & $0.053\pm0.004$\\
NGC~4194          & $ 257\pm  26$ & $ 193\pm  13$ & $    \cdots $ & $ 11.11\pm  0.26$ & $      \cdots   $ & $43.54\pm0.05$ & $43.41\pm0.03$ & $    \cdots  $ & $41.18\pm0.01$ & $    \cdots   $\\
NGC~4818          & $1191\pm  54$ & $ 181\pm  35$ & $    \cdots $ & $ 10.43\pm  0.33$ & $      \cdots   $ & $43.46\pm0.02$ & $42.64\pm0.09$ & $    \cdots  $ & $40.40\pm0.01$ & $    \cdots   $\\
NGC~520           & $1052\pm  42$ & $ 206\pm  19$ & $    \cdots $ & $ 15.10\pm  0.31$ & $      \cdots   $ & $44.07\pm0.02$ & $43.36\pm0.04$ & $    \cdots  $ & $41.23\pm0.01$ & $    \cdots   $\\
NGC~6285          & $ 120\pm  13$ & $  26\pm   6$ & $    \cdots $ & $  1.86\pm  0.07$ & $      \cdots   $ & $43.93\pm0.05$ & $43.25\pm0.11$ & $    \cdots  $ & $41.12\pm0.02$ & $    \cdots   $\\
NGC~838           & $ 370\pm  28$ & $ 137\pm  22$ & $    \cdots $ & $ 14.55\pm  0.15$ & $      \cdots   $ & $44.07\pm0.03$ & $43.64\pm0.08$ & $    \cdots  $ & $41.67\pm0.01$ & $    \cdots   $\\
ZW~453.062        & $ 245\pm   6$ & $    \cdots $ & $  21\pm   6$ & $  2.47\pm  0.16$ & $  0.32\pm  0.07$ & $44.49\pm0.01$ & $    \cdots  $ & $43.42\pm0.14$ & $41.49\pm0.03$ & $0.031\pm0.008$\\
\enddata
\tablecomments{Table~3 summarizes the observed properties including continuum fluxes and luminosities of the 22 sources obtained from the {\it AKARI} spectroscopy. 
(1) object name: 
(2), (3), (4) flux of the stellar, \textsc{Hii}, and AGN blackbody component in units of $10^{-12}$~erg cm$^{-2}$ s$^{-1}$, respectively: 
(5) 3.3~$\mu$m PAH flux in units of $10^{-13}$~erg cm$^{-2}$ s$^{-1}$:
(6) optical thickness of H$_2$O ice:
(7), (8), (9) logarithmic luminosity of the stellar, \textsc{Hii}, and dust-torus blackbody component in units of erg s$^{-1}$, respectively:
(10) logarithmic luminosity of the 3.3~$\mu$m PAH emission in units of erg s$^{-1}$:
(11) ratio of the AGN blackbody luminosity to the total infrared luminosity. 
}\\
\end{deluxetable*}
\end{turnpage}

 \subsection{Total model spectra}

We thus construct a spectral model consisting of the multiple continuum
components over which the emission/absorption lines are superposed. 
For the continuum we consider three blackbody components, 
(1) a direct stellar component $f_{\rm BB}^{\rm (stellar)}$, 
(2) a dust component from AGN with a temperature below 800~K $f_{\rm  BB}^{\rm (dust)}$, and 
(3) a very hot (800~K--1200~K) dust component from \textsc{Hii}
regions $f_{\rm BB}^{(\textsc{Hii})}$.
Each blackbody component has two parameters, 
normalization ($C^{\rm (stellar)}, C^{\rm (dust)}, C^{(\textsc{Hii})}$)
and temperature ($T^{\rm (stellar)}, T^{\rm (dust)}, T^{(\textsc{Hii})}$),
which, except for $T^{\rm (stellar)}$, are left as free parameters in the spectral fit.

{Previous studies suggest that emission from warm dust may be 
partly absorbed by outer cold dust \citep[e.g.,][]{ris10,ima10}.
To take this into account, we apply an extinction correction 
by incorporating $\tau(\lambda)$ into the blackbody components.
We adopt the dust extinction index of $\beta=1.75$
\citep{dra89,hun02} as
\begin{align}
\tau(\lambda) = \tau_0 \left(\frac{\lambda}{\mu{\rm m}}\right)^{-1.75}.
 \end{align}
The normalization value $\tau_0$ can be estimated from the optical thickness of 3.1~$\mu$m ice ($\tau_{\rm ice}$),
which is calculated by using the deconvolved continuum from the dust
($f_{\rm BB}^{(\rm dust)}$). \cite{ima03} estimated the relationship
between the dust extinction $A_{V}$ and $\tau_{\rm ice}$ with
\begin{align}
\tau_{\rm ice} = 0.06 A_V (1+f),
\end{align}
where $f$ is the fraction of dust that is covered with an ice mantle. 
The maximum value of $f$ is
derived to be 0.3 in the core of U/LIRGs in \cite{ima03}. 
From the relation between the optical thickness and extinction 
with $\tau_V = A_V / 1.08$ and Eq.~(2) with $f=0.3$,
\begin{align}
\tau_V = \frac{\tau_{\rm ice}}{0.06\times1.3\times1.08} = 11.87 \tau_{\rm ice}
\end{align}
is derived. By combining Eq.~(1) and (3), at the $V$ band (0.55~$\mu$m),
\begin{align}
\tau(\lambda=0.55~\mu{\rm m}) &= \tau_0 \times 0.55^{-1.75} = 11.87\tau_{\rm ice} \\
\iff \tau_0 &= 4.3\tau_{\rm ice}
\end{align}
Accordingly,
we model ths observed component as 
$f_{\rm BB obs}^{\rm (dust)} =  f_{\rm BB}^{\rm (dust)}(\lambda) \times e^{-\tau(\lambda)}$,
where $\tau(\lambda) = 4.3 \times \tau_{\rm ice} \times (\lambda/\mu {\rm m})^{-1.75}$.

First, we adopt only two continuum components of direct stellar
emission and AGN/SB dust emission to fit the observed spectra, 
because the contribution of dust emission from \textsc{Hii} regions is
not always required from the data. 
This model (2 component model) is written as:
 \begin{align}
 f_{\rm model}(\lambda) = f_{\rm BB}^{\rm (stellar)}(\lambda) + f_{\rm BBobs}^{\rm (dust)}(\lambda) + \sum_{i} f_i^{\rm (line)}(\lambda),
 \end{align}
where $f_i^{\rm (line)}(\lambda)$ represents each line's profile as
described in Section~4.3. In some cases, mainly for sources showing 
little contribution from $f_{\rm BB}^{\rm (dust)}$, the 2-component model 
given by equation (6) does not fit the data well. Then, we add 
the $f_{\rm BB}^{\rm \textsc{(Hii)}}$ component to the above model:
  \begin{align}
\hspace{-0.5cm} f_{\rm model}(\lambda) = f_{\rm BB}^{\rm (stellar)}(\lambda) + f_{\rm BB}^{\rm \textsc{(Hii)}}(\lambda) + f_{\rm BBobs}^{\rm (dust)}(\lambda) + \sum_{i} f_i^{\rm (line)}(\lambda).
 \end{align}
We adopt this model (3 component model) only when the fitting result
is significantly improved from the 2-component model with a $>95$\%
confidence level on the basis of an F-test. 
The fitting model we adopt
for each source is summarized in Table~1. 
As shown in Figure~2,  all sources that require the three
components are galaxies showing blue contina
in their spectra (e.g., NGC~4818). 
As many authors \citep[e.g.,][]{ris10, ima10} suggested, 
these galaxies do not show buried AGN signs but are more likely
 normal SB galaxies.
This is consistent with our assumption that the extra hot ($T\sim10^3$~K) component should
originate from} \textsc{Hii} regions, not from the AGN torus.

\begin{figure*}[tbp]
\begin{center}
\includegraphics[angle=0,scale=0.2]{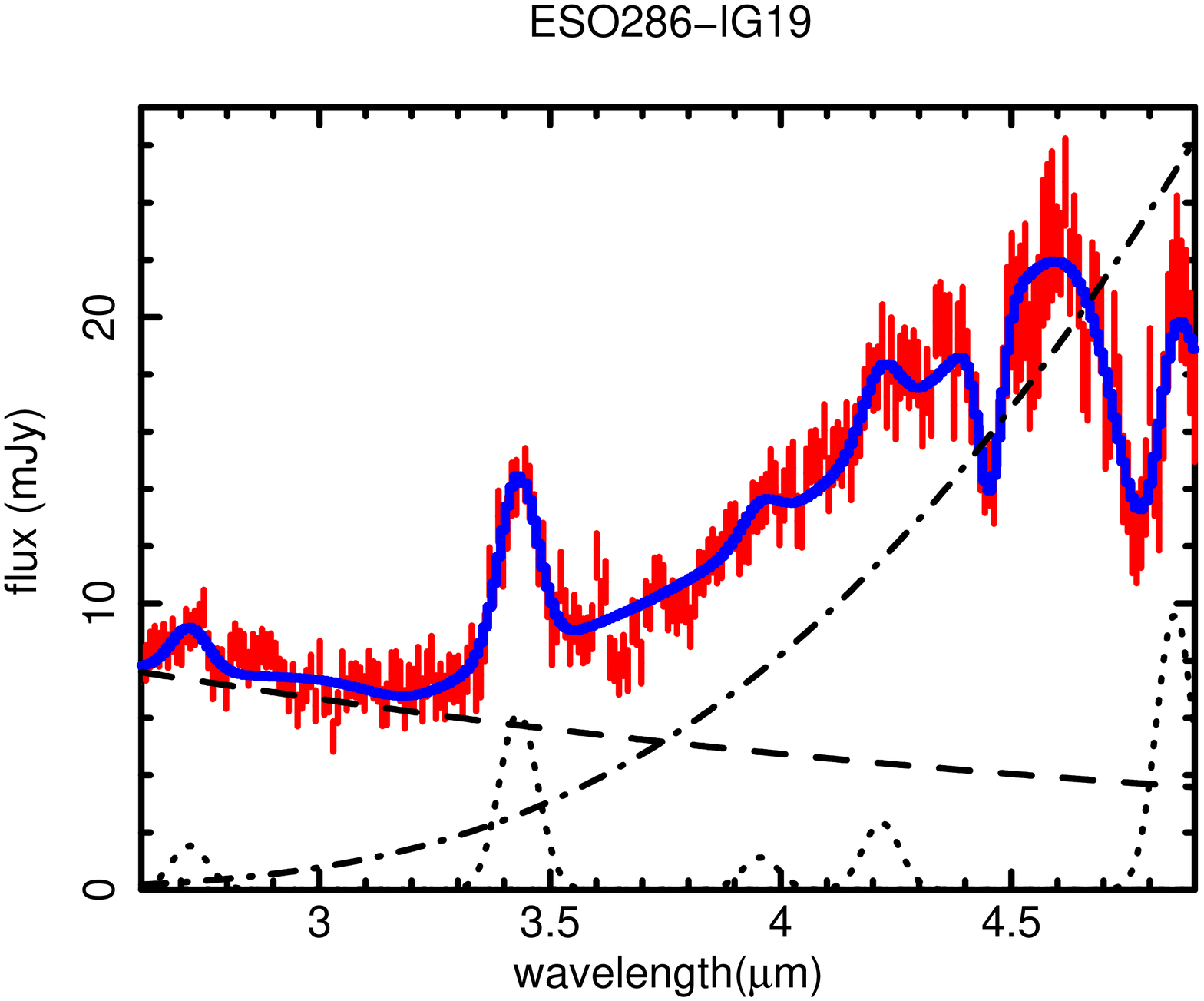}
\includegraphics[angle=0,scale=0.2]{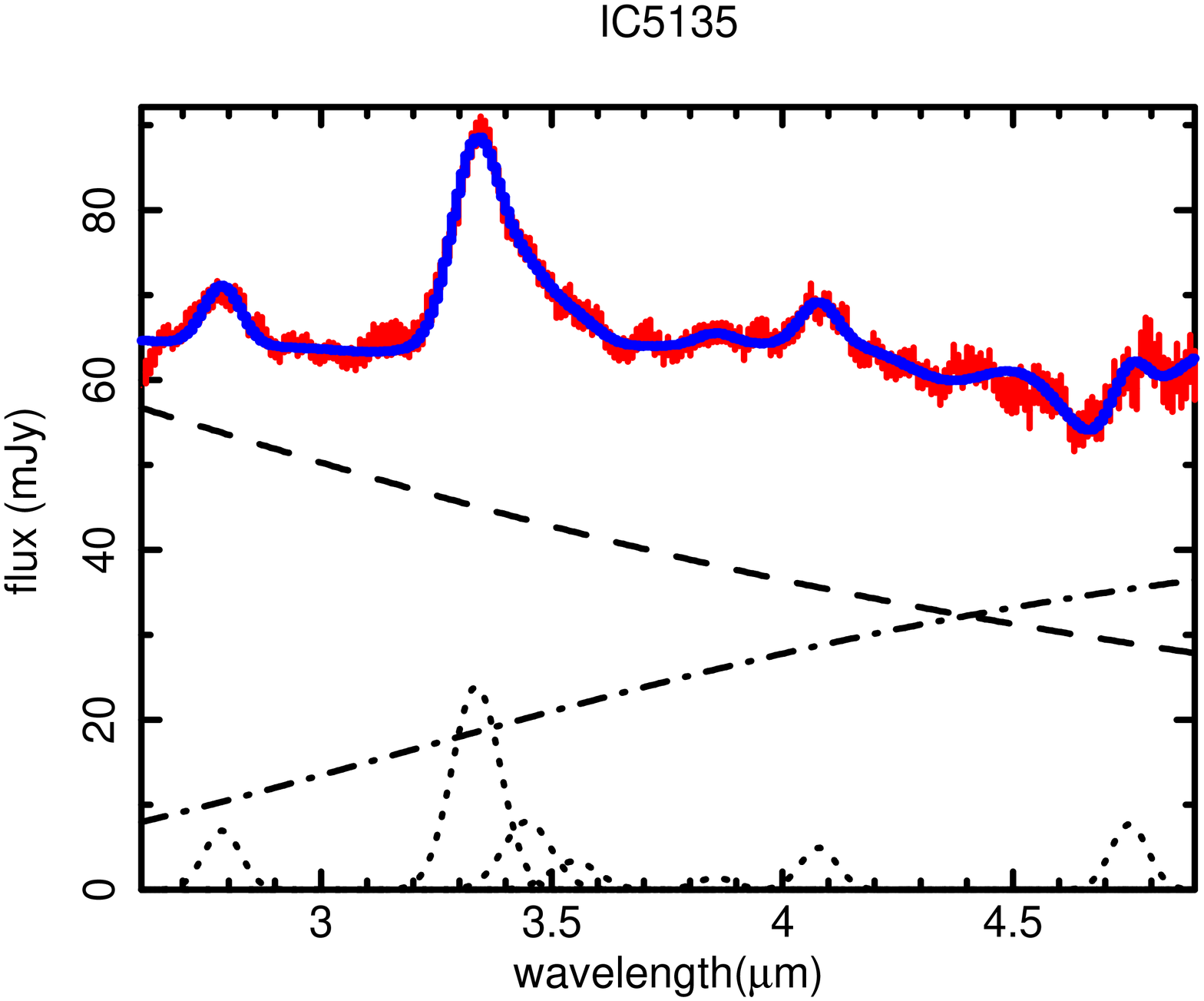}
\includegraphics[angle=0,scale=0.2]{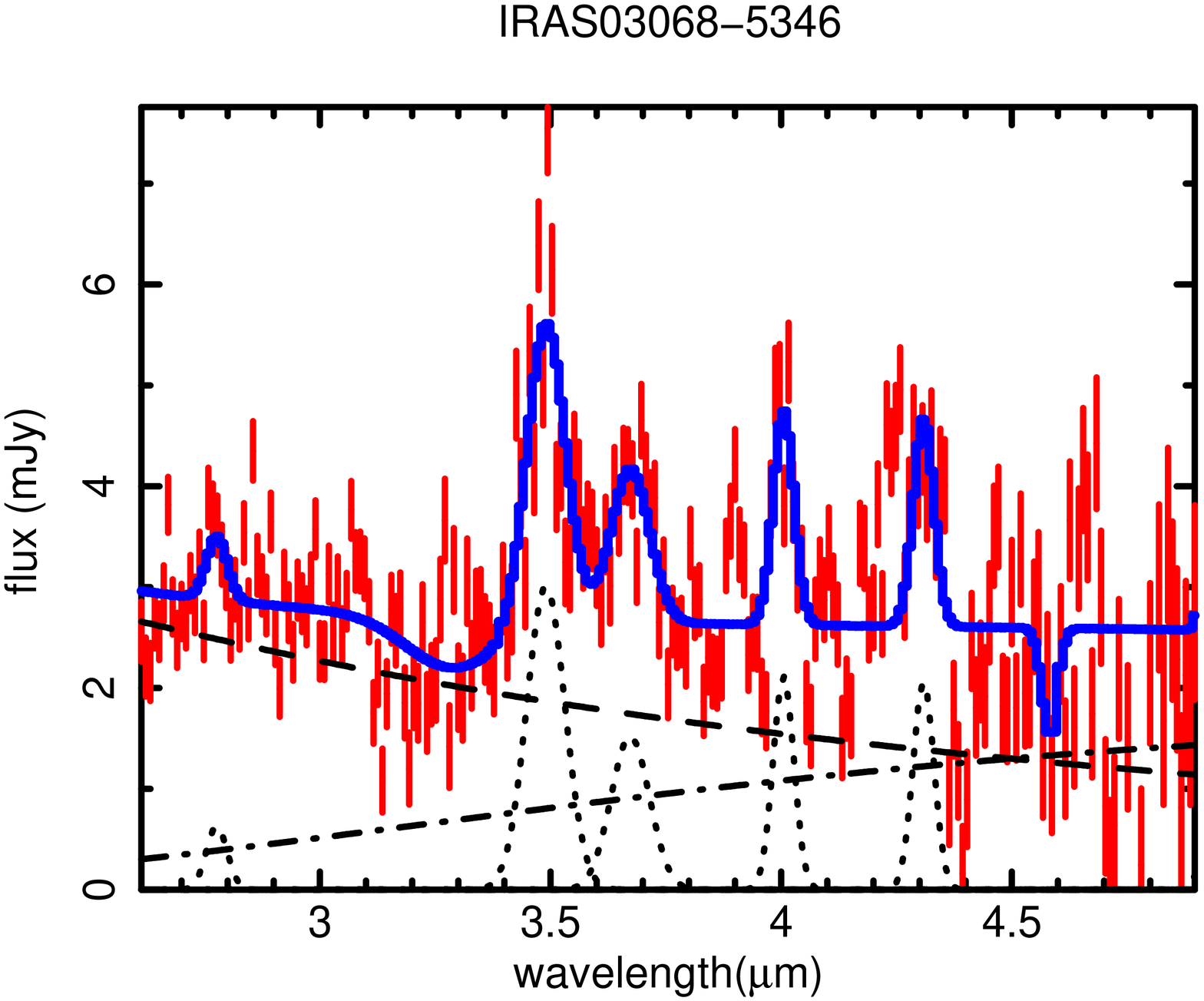}
\includegraphics[angle=0,scale=0.2]{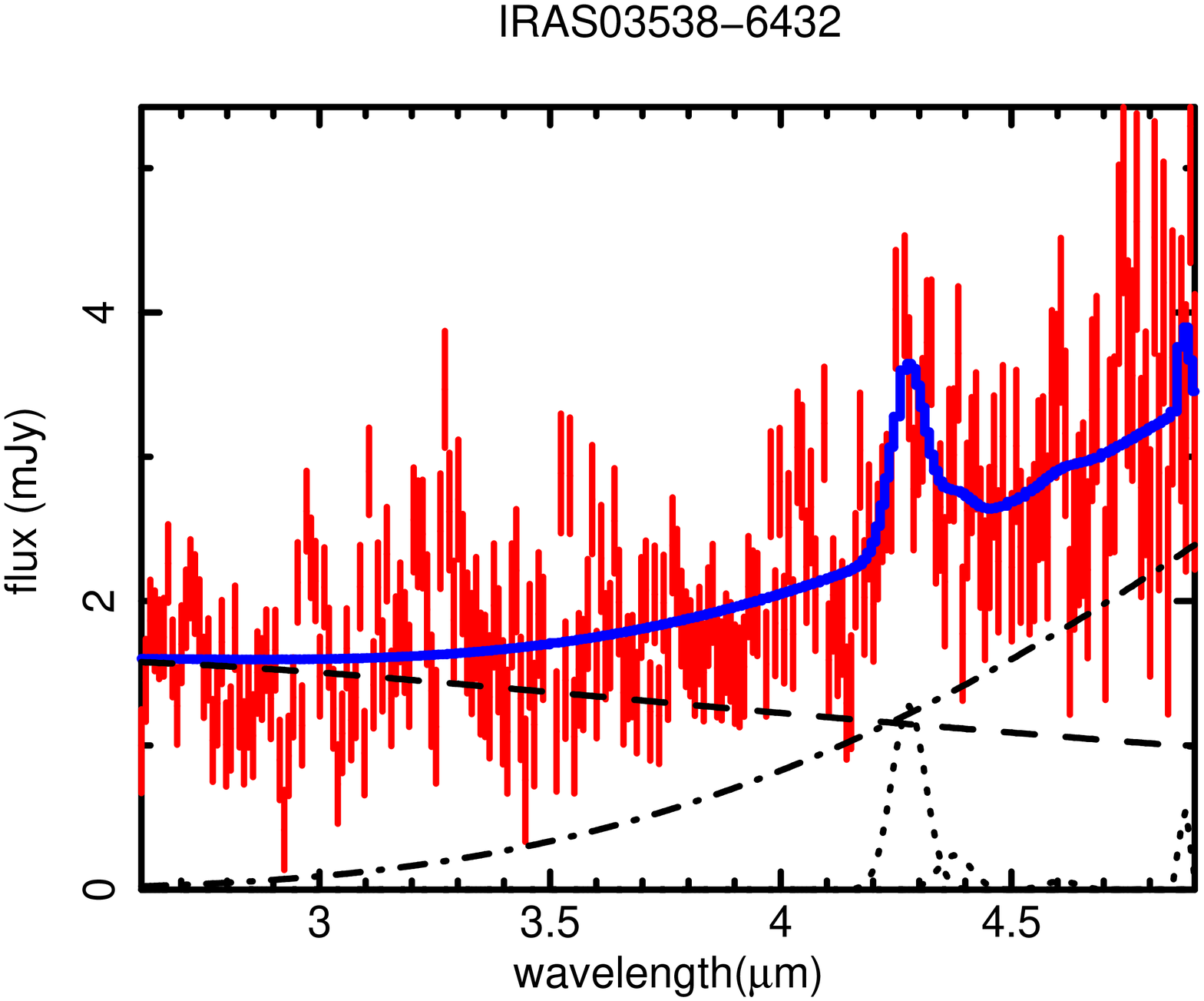}
\includegraphics[angle=0,scale=0.2]{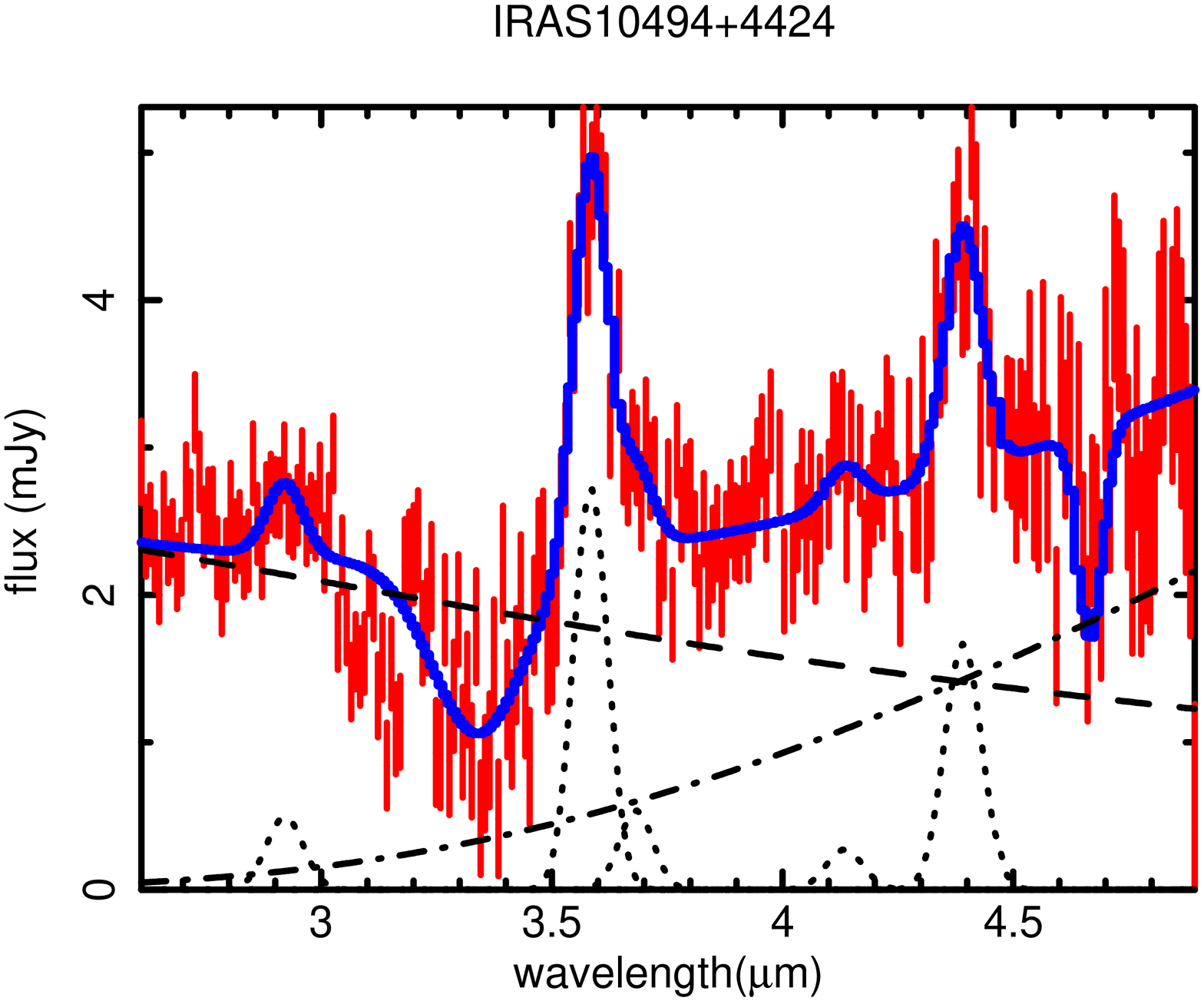}
\includegraphics[angle=0,scale=0.2]{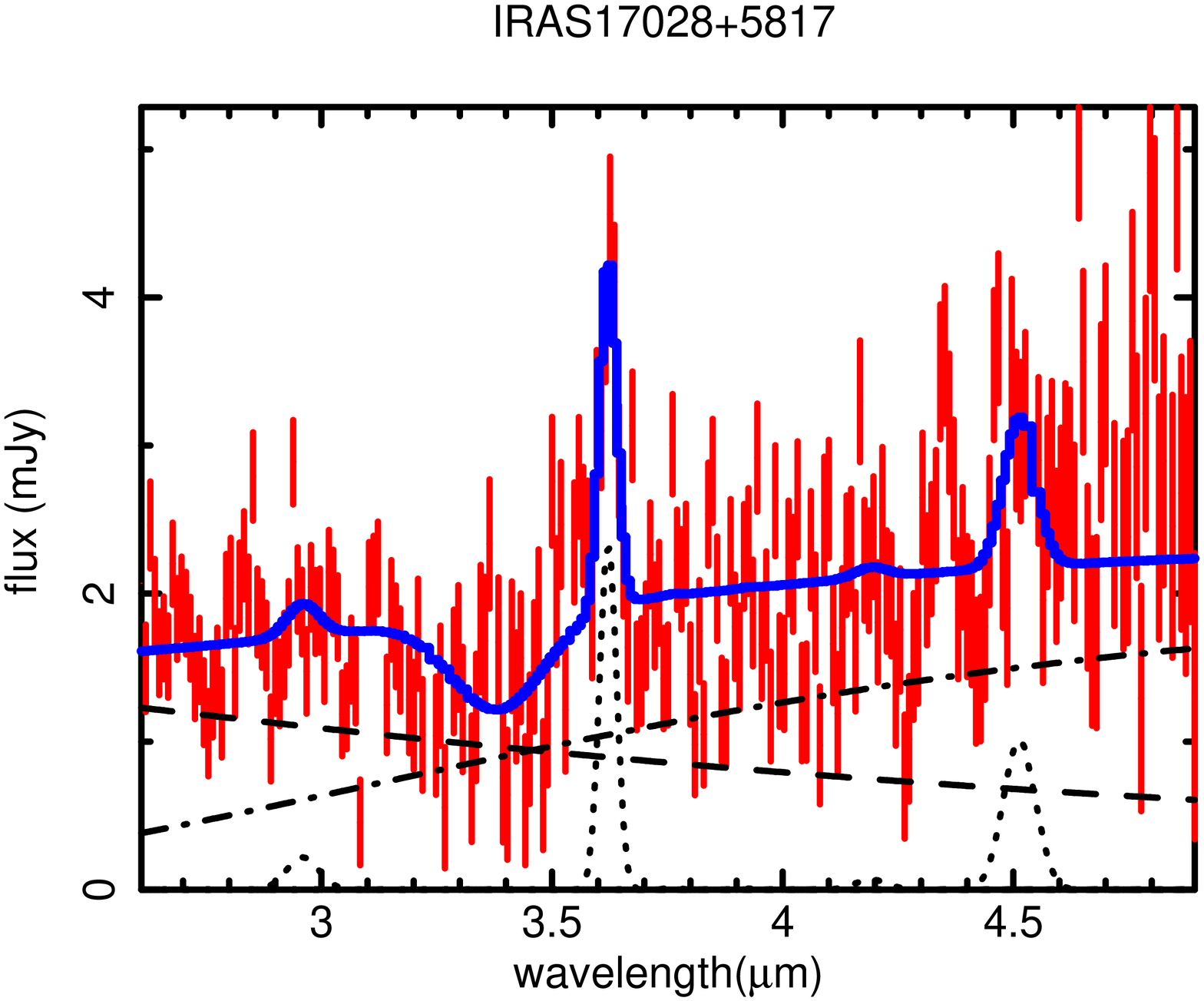}
\includegraphics[angle=0,scale=0.2]{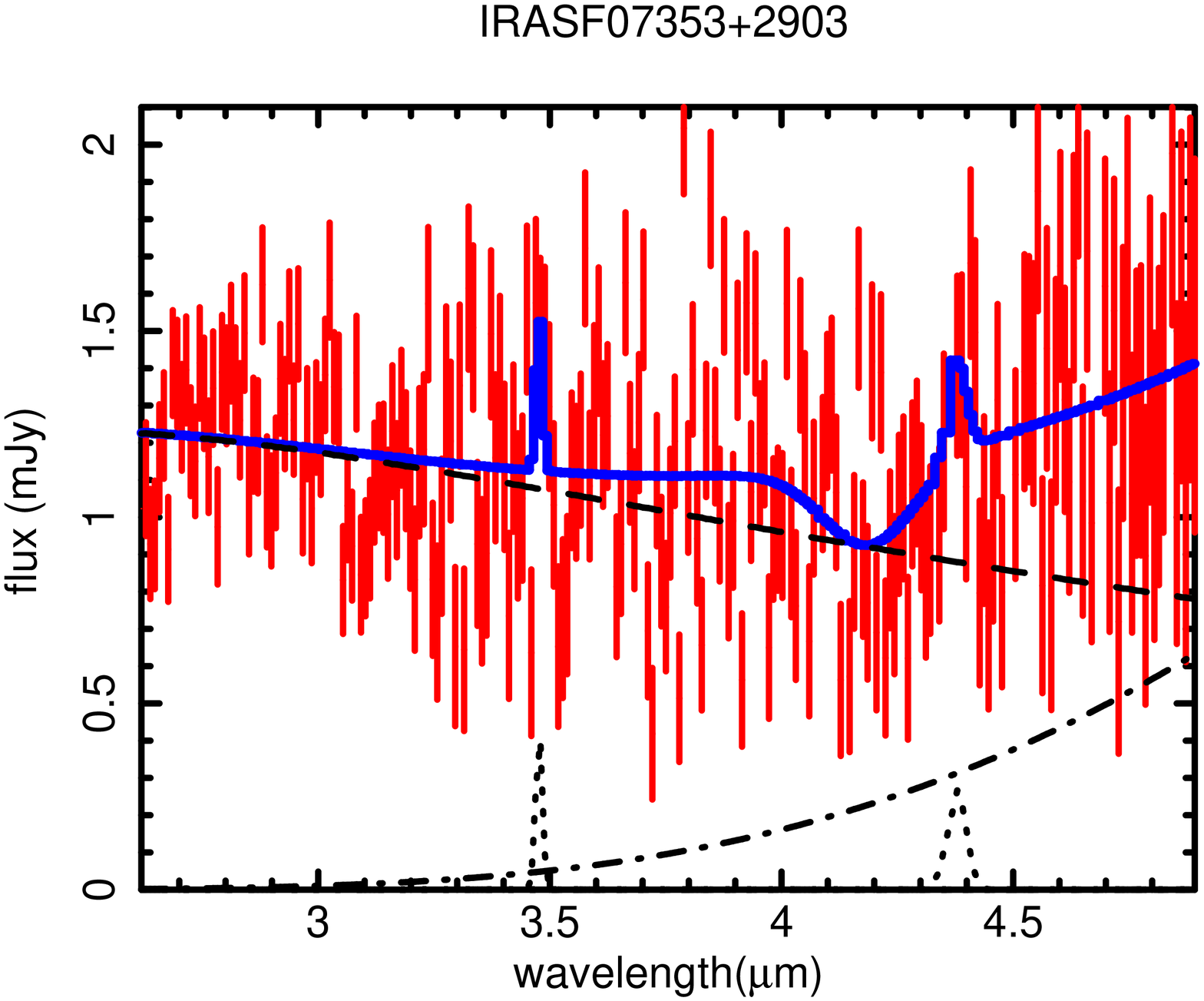}
\includegraphics[angle=0,scale=0.2]{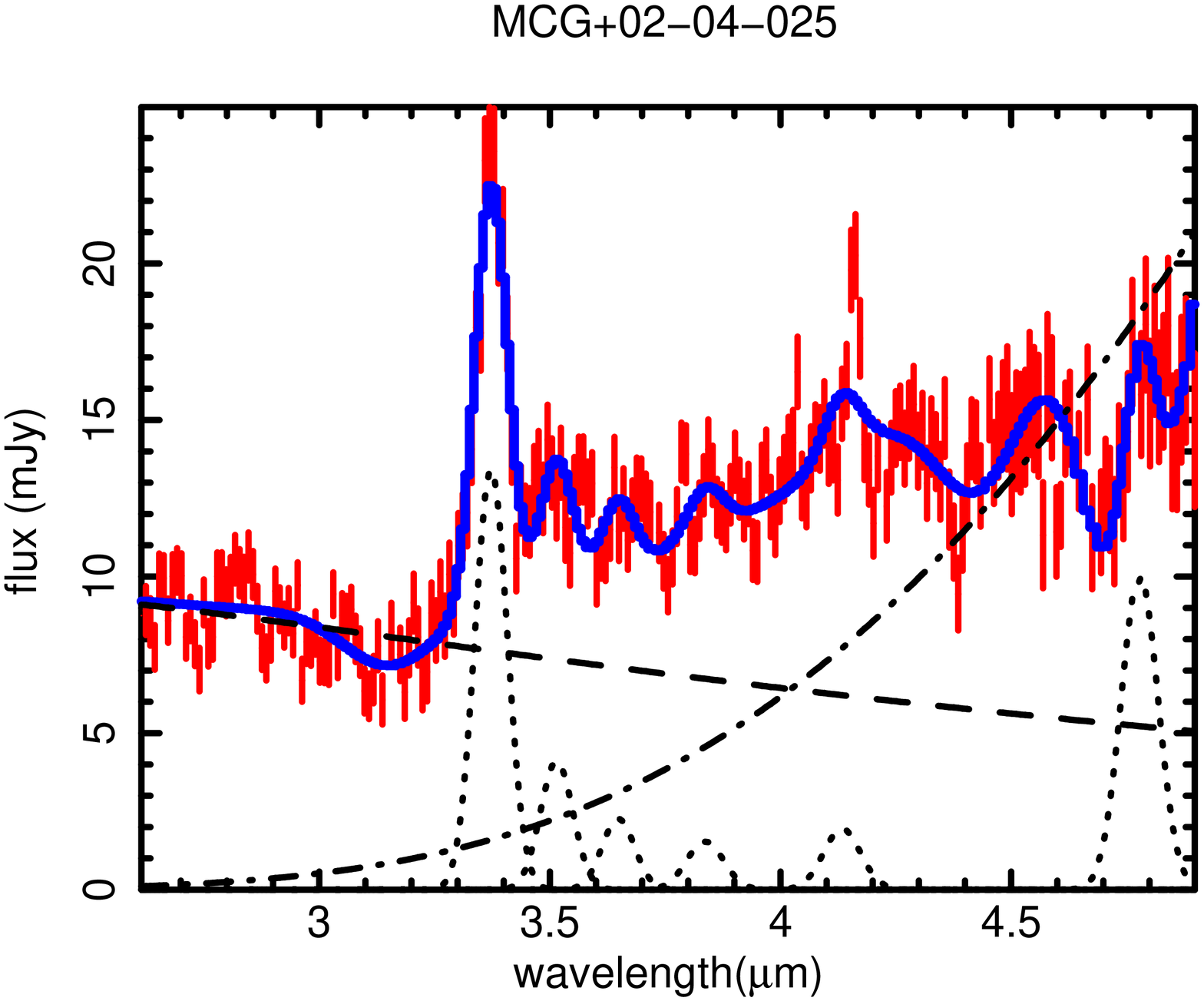}
\includegraphics[angle=0,scale=0.2]{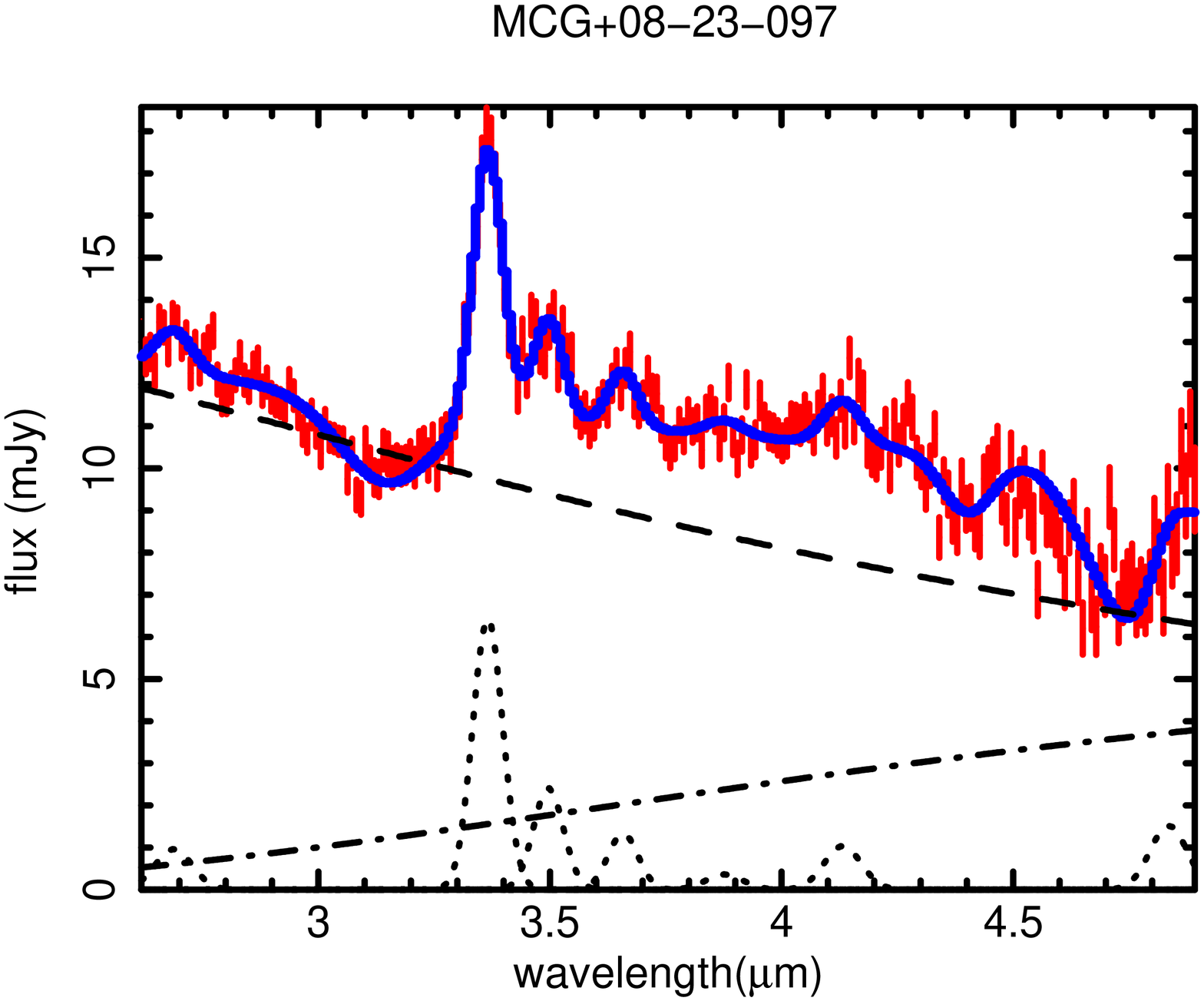}
\includegraphics[angle=0,scale=0.2]{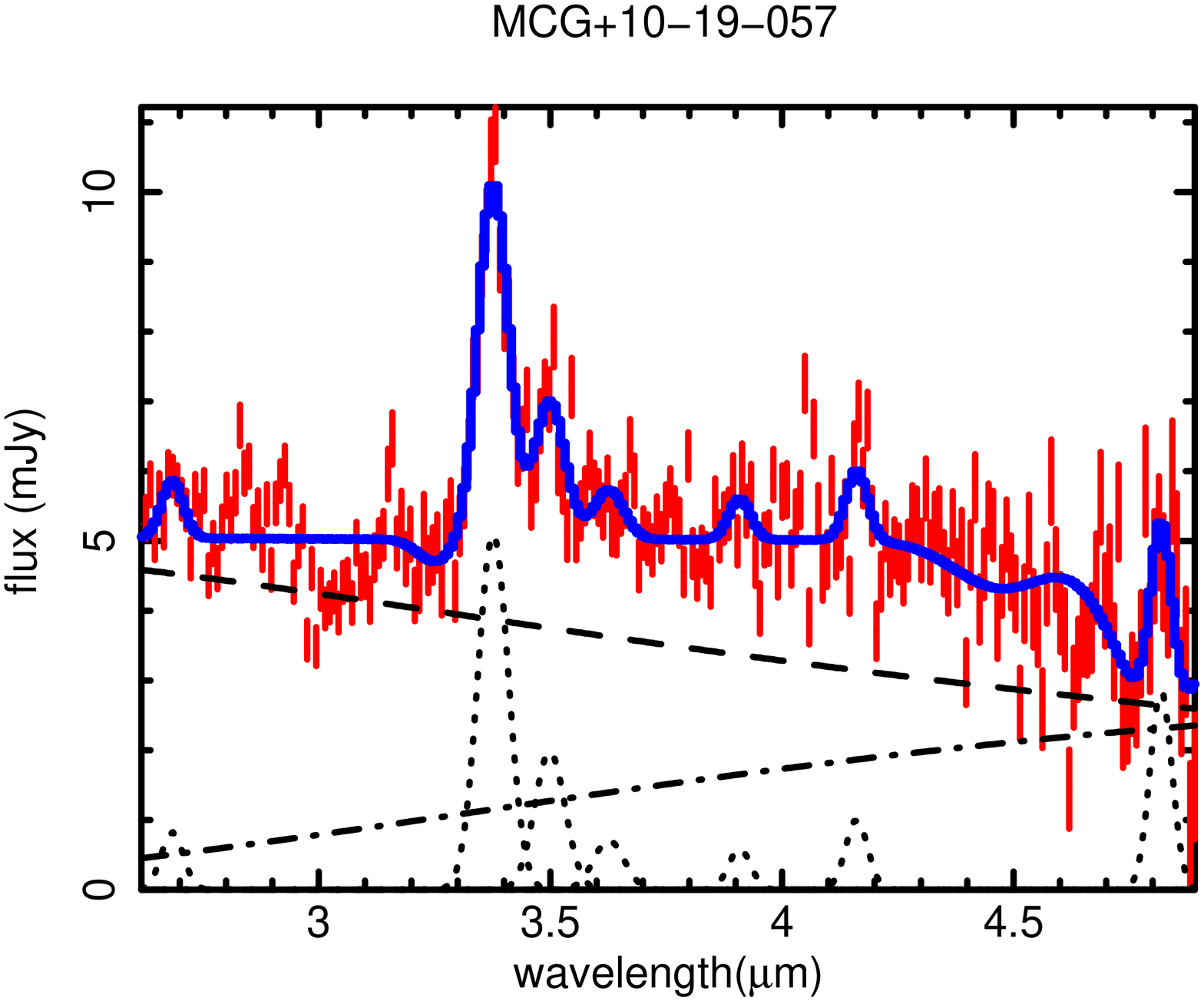}
\includegraphics[angle=0,scale=0.2]{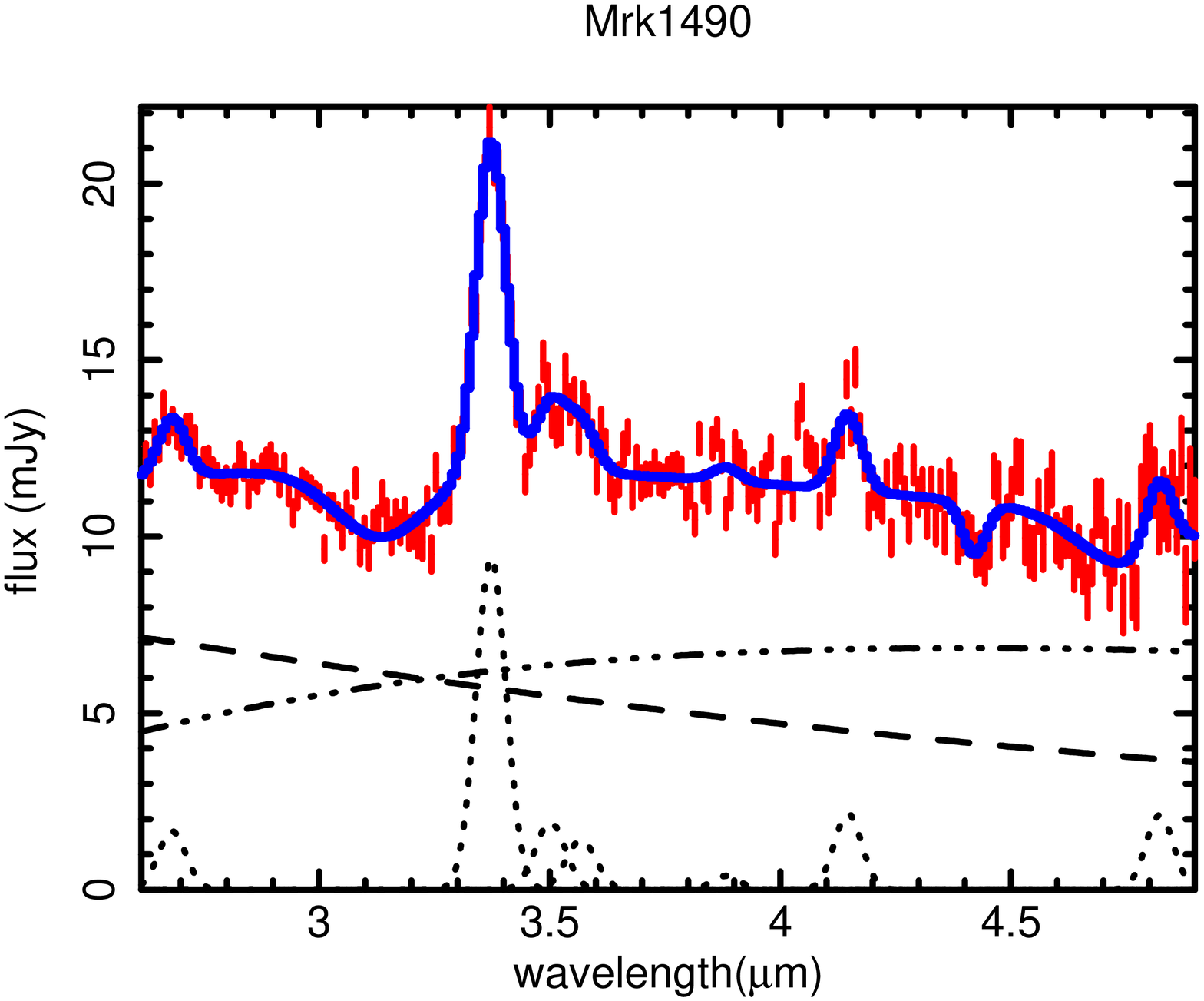}
\includegraphics[angle=0,scale=0.2]{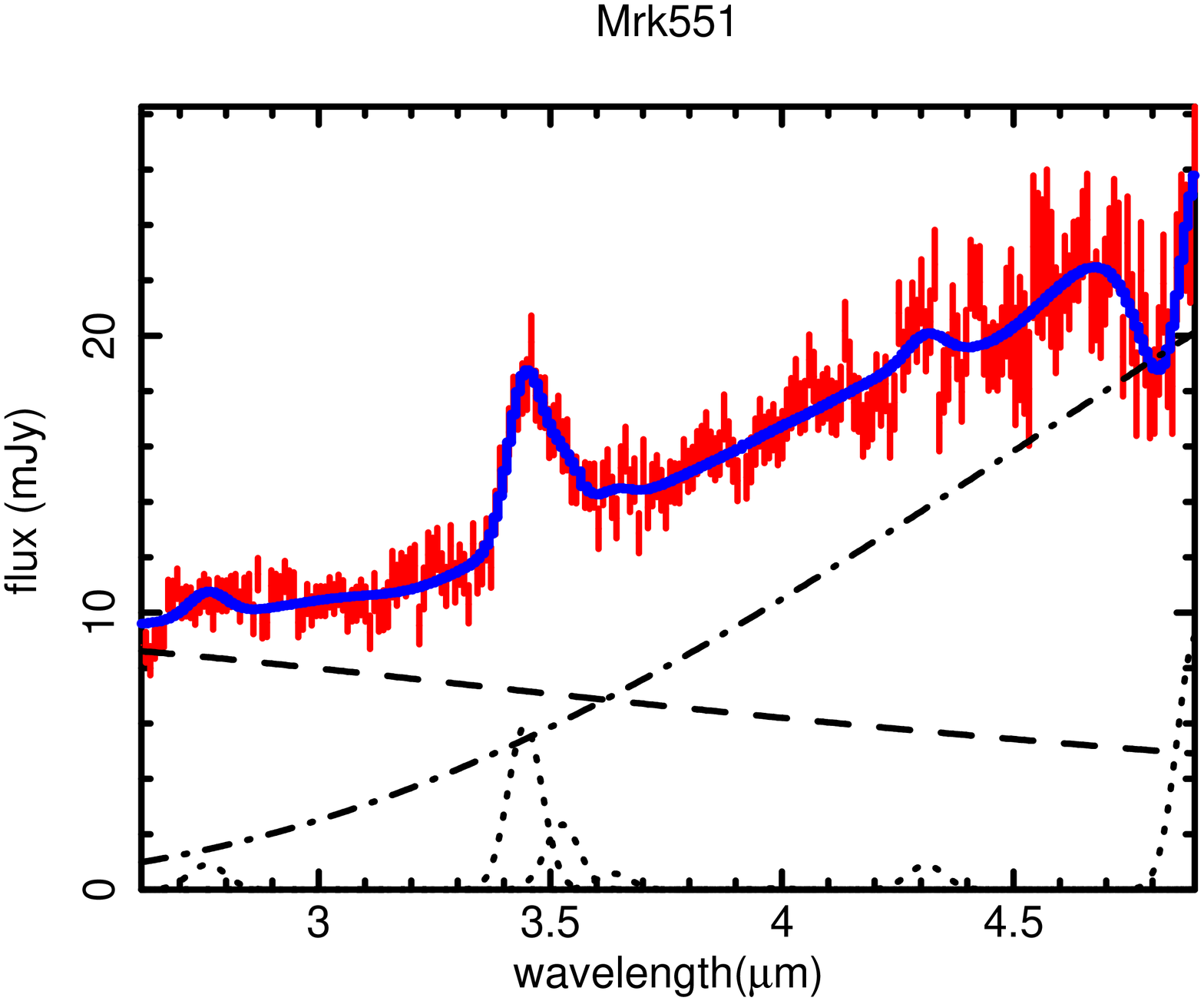}
\\
\caption{
\textit{AKARI} 2.5--5.0~$\mu$m spectra (red dots with error bars) of
the 22 infrared galaxies overplotted with the best-fit models (blue
solid curve). The black dashed, dashed-three-dot, dashed-dot curves
represent the stellar, \textsc{Hii}, and AGN/SB dust components,
respectively.  Black dotted curves represent emission lines.
\label{fig-2}}
\end{center}
\end{figure*}

\addtocounter{figure}{-1}[tbp]
\begin{figure*}
\begin{center}
\includegraphics[angle=0,scale=0.2]{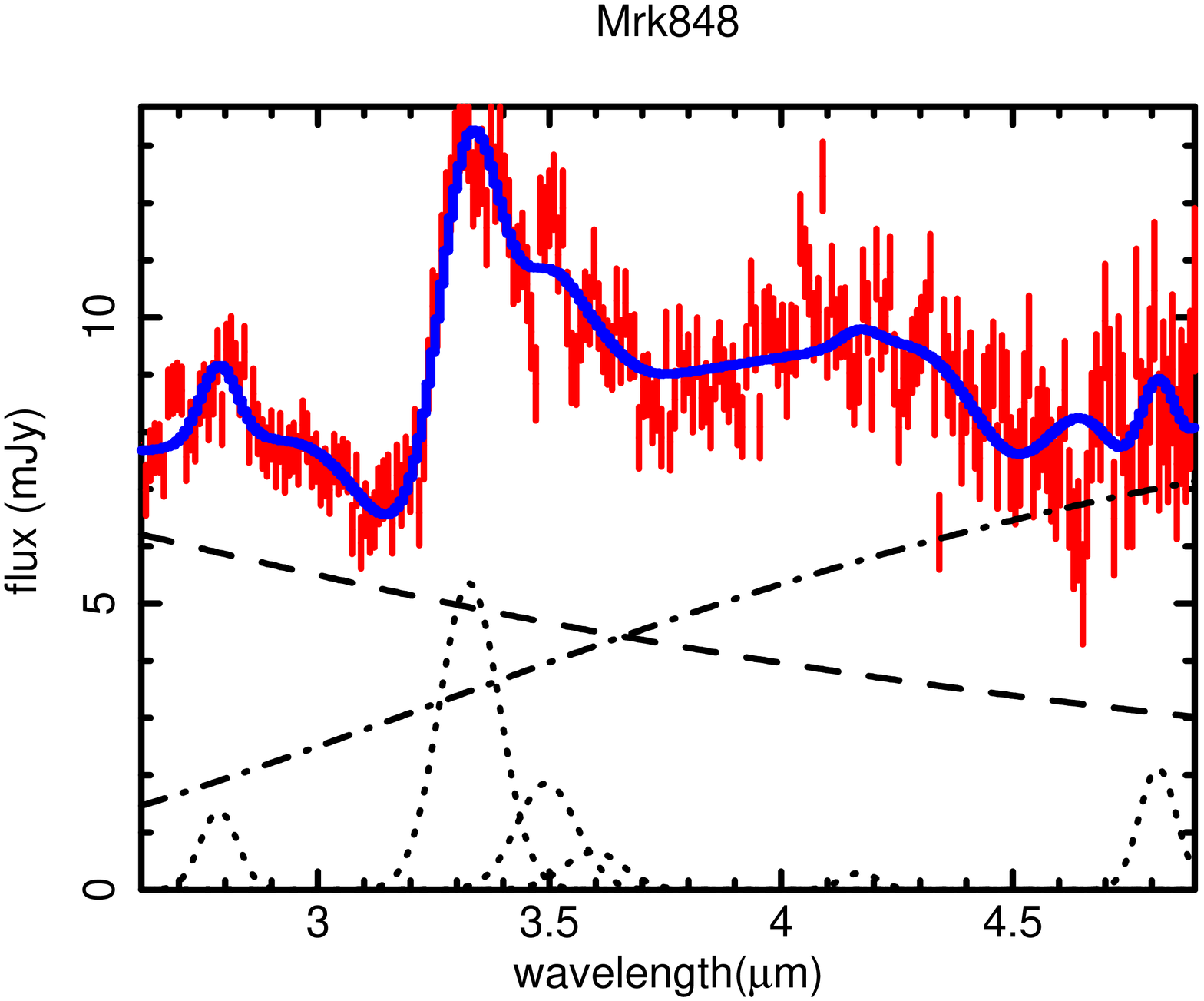}
\includegraphics[angle=0,scale=0.2]{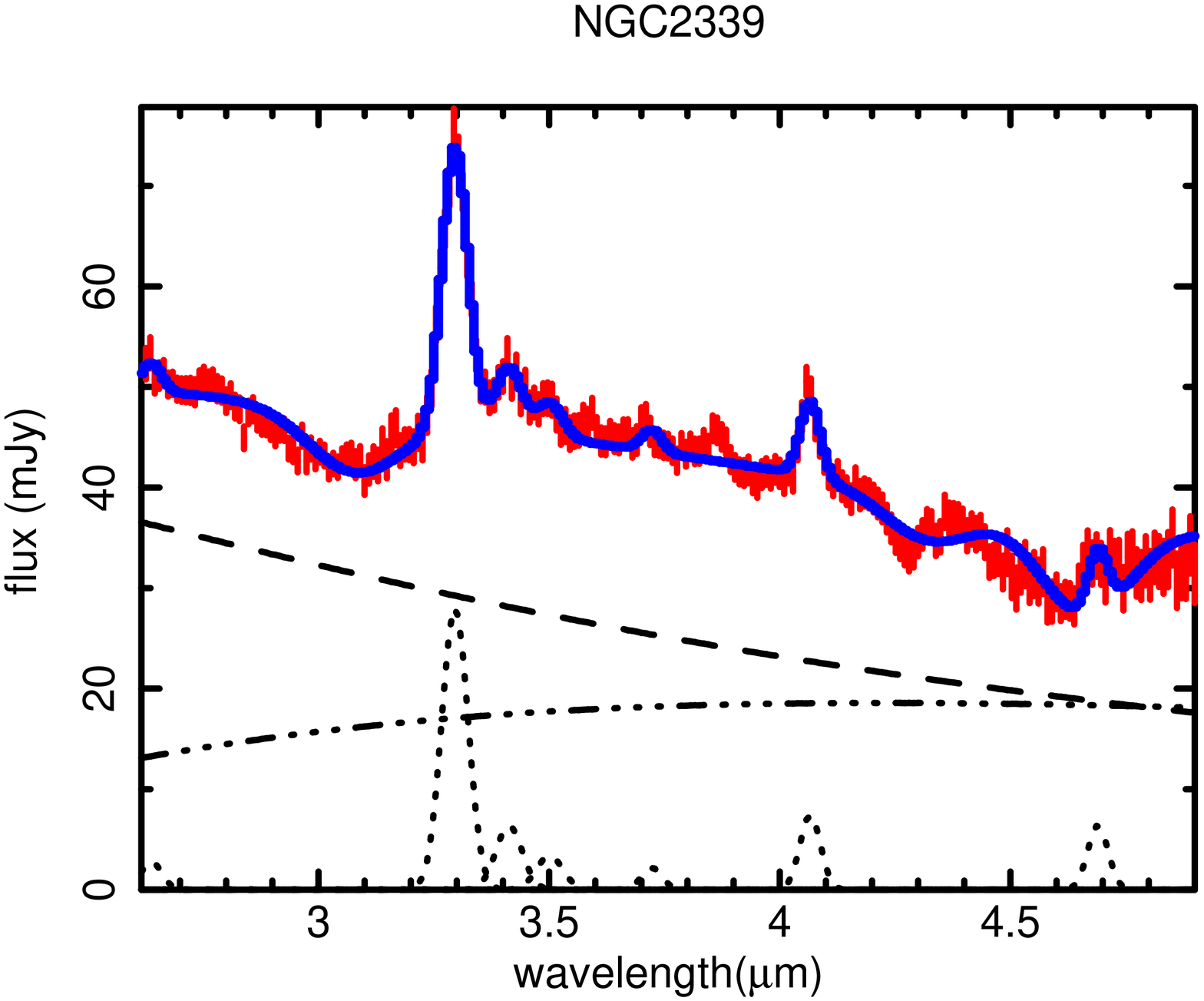}
\includegraphics[angle=0,scale=0.2]{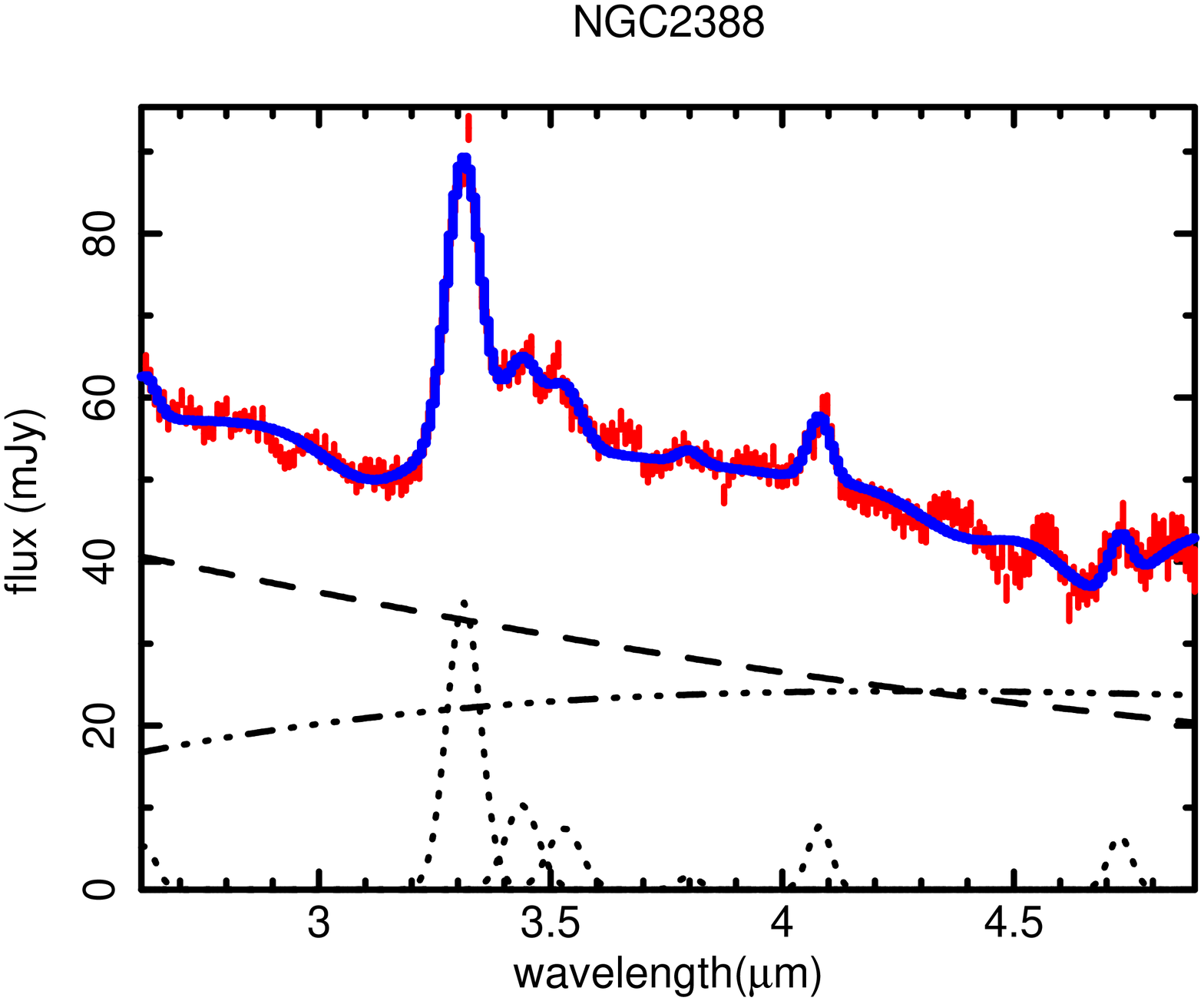}
\includegraphics[angle=0,scale=0.2]{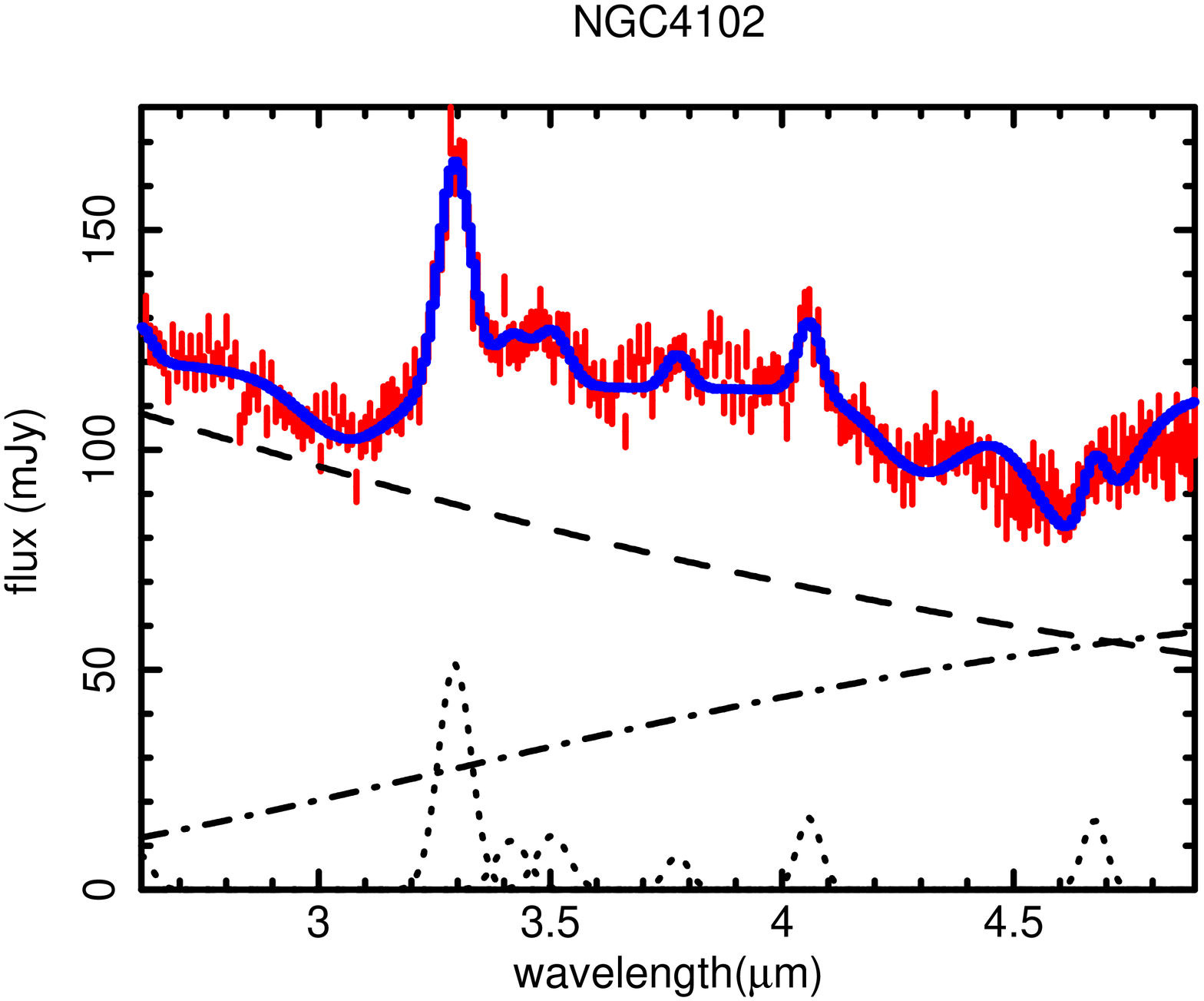}
\includegraphics[angle=0,scale=0.2]{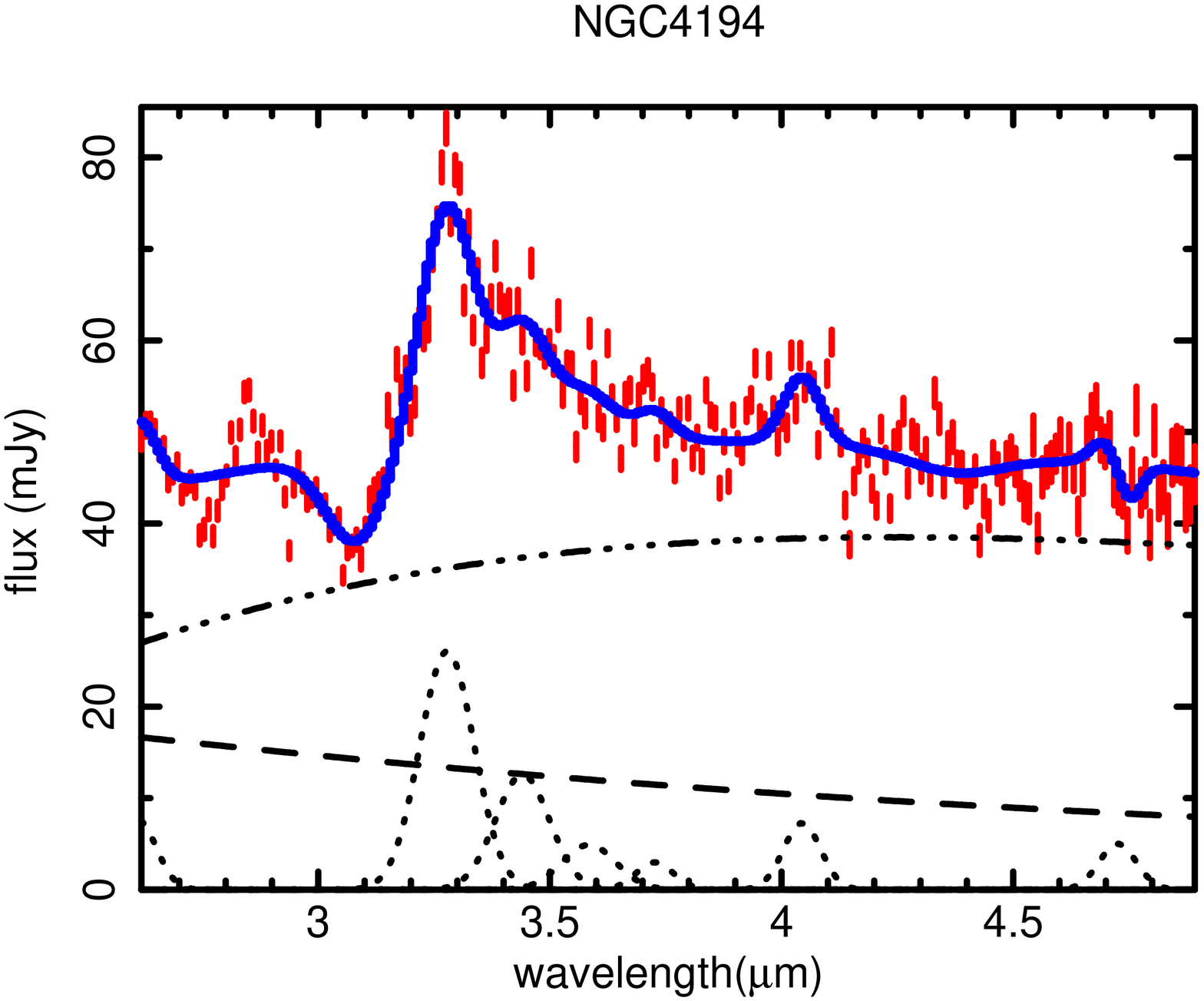}
\includegraphics[angle=0,scale=0.2]{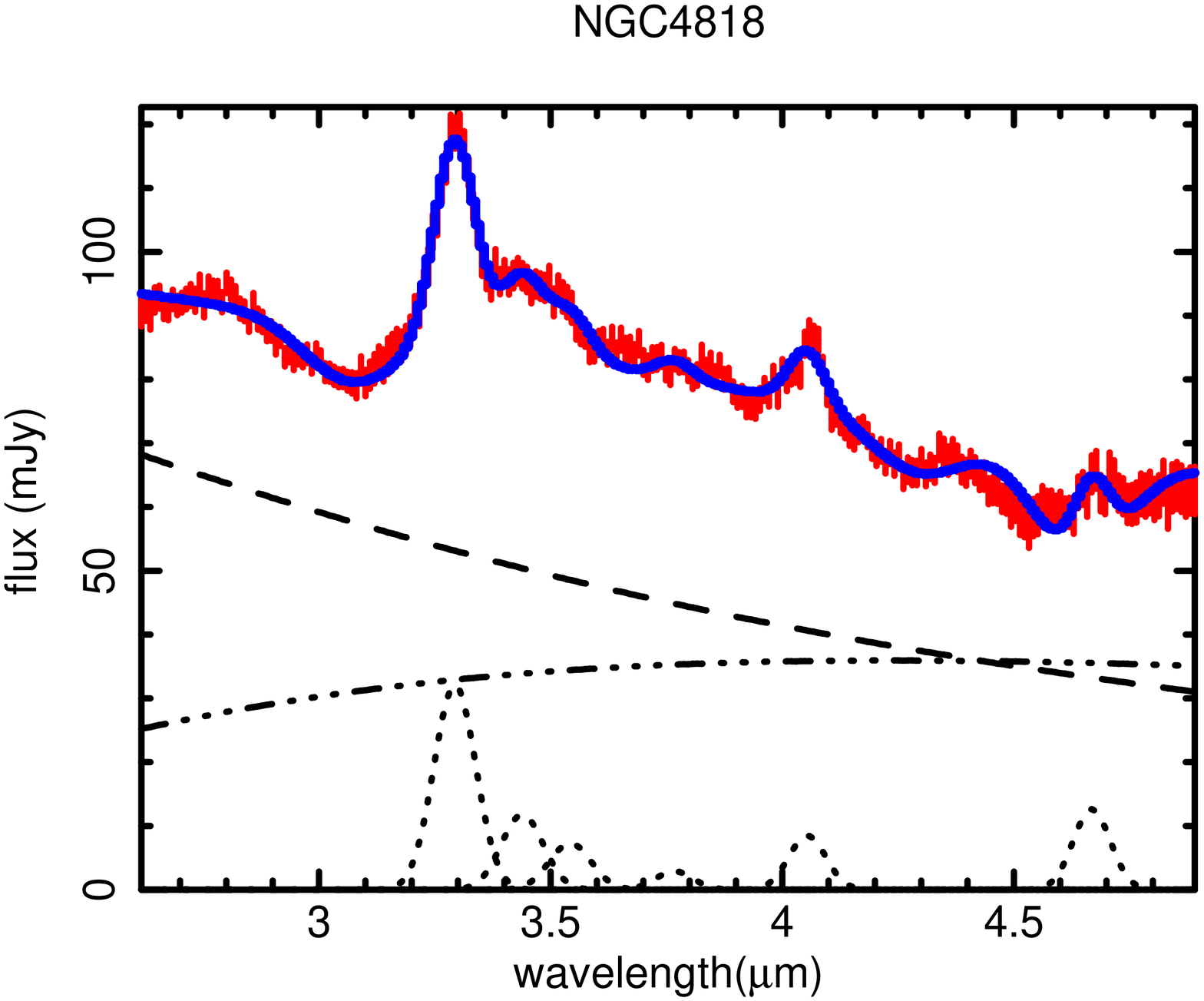}
\includegraphics[angle=0,scale=0.2]{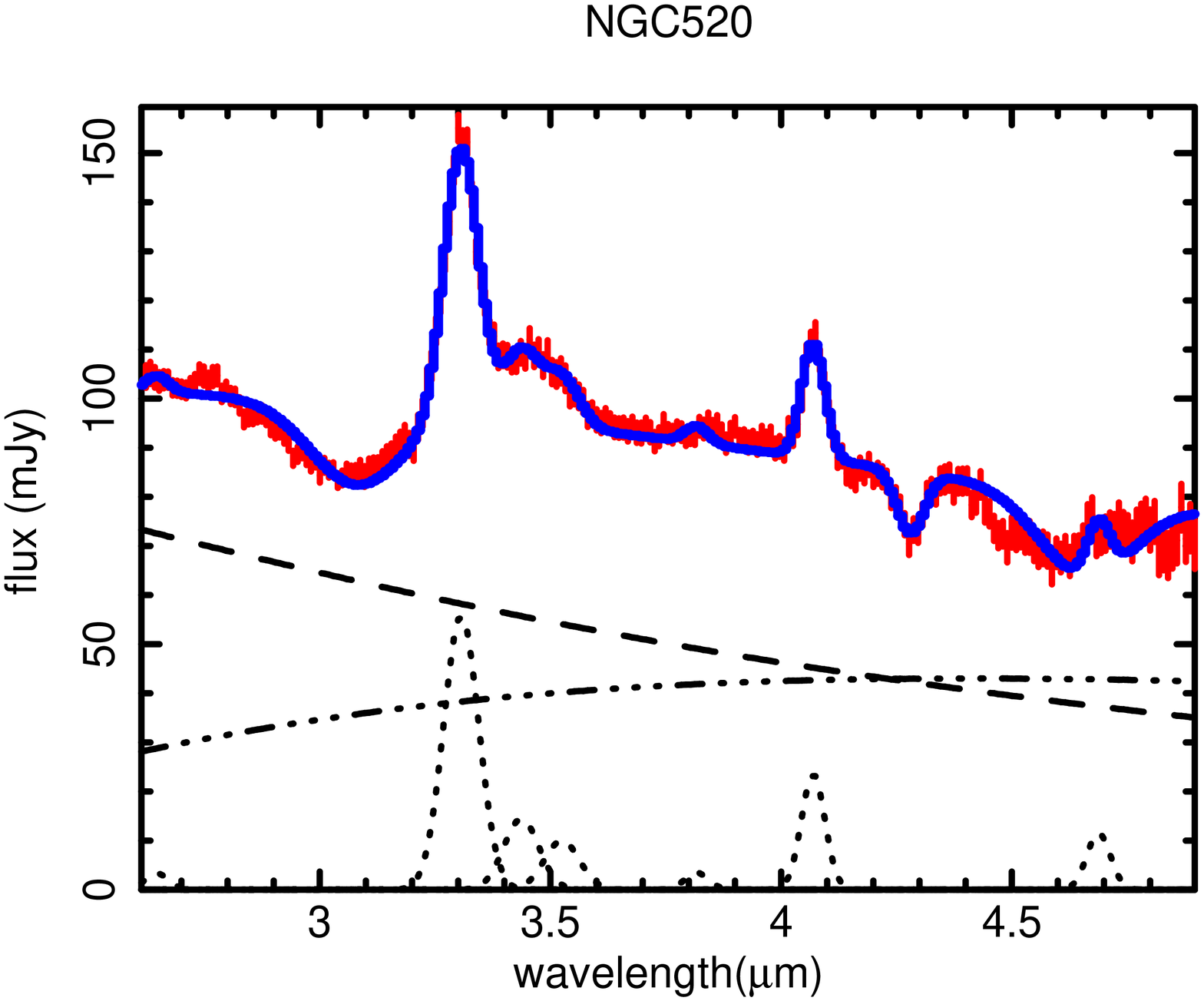}
\includegraphics[angle=0,scale=0.2]{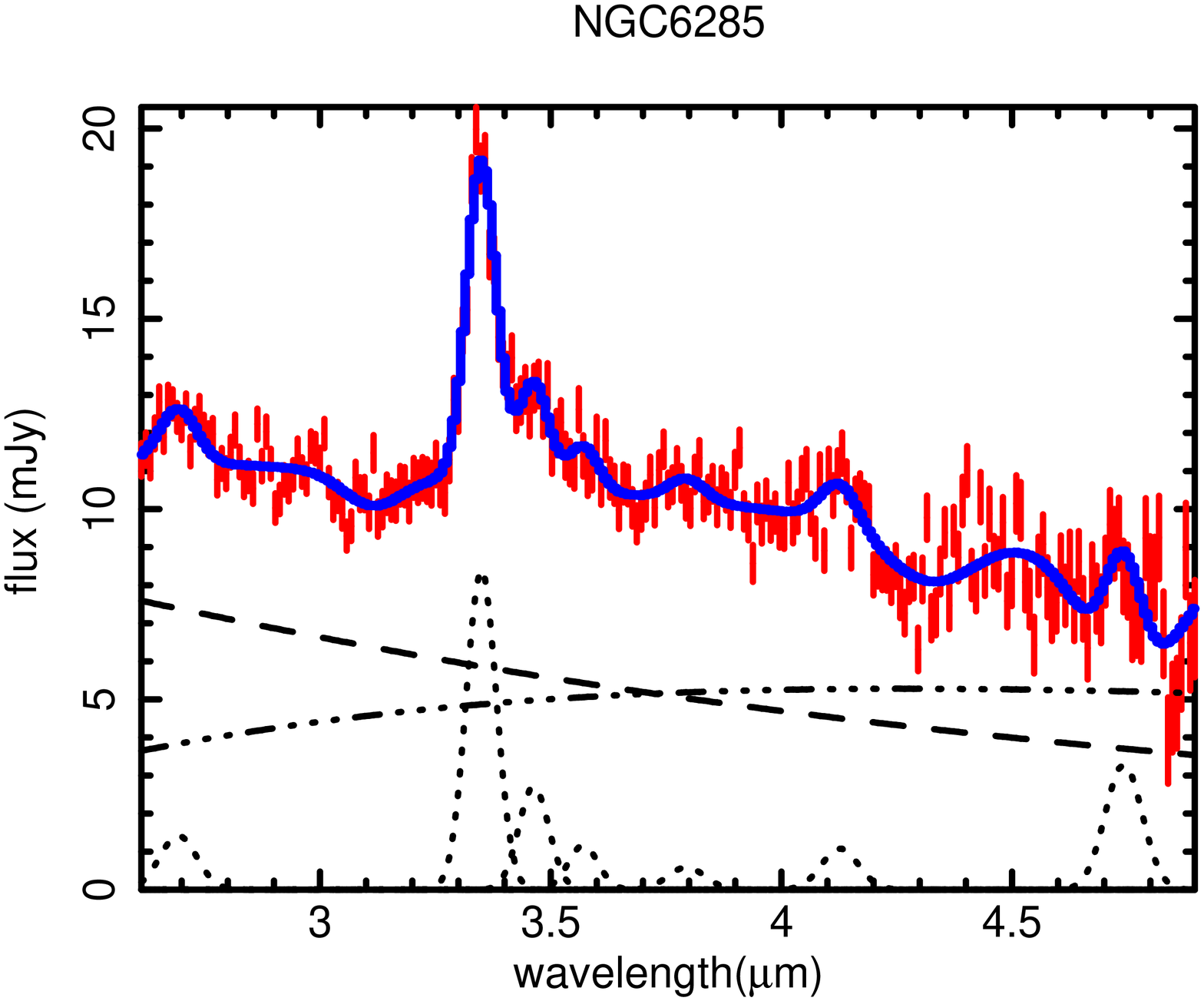}
\includegraphics[angle=0,scale=0.2]{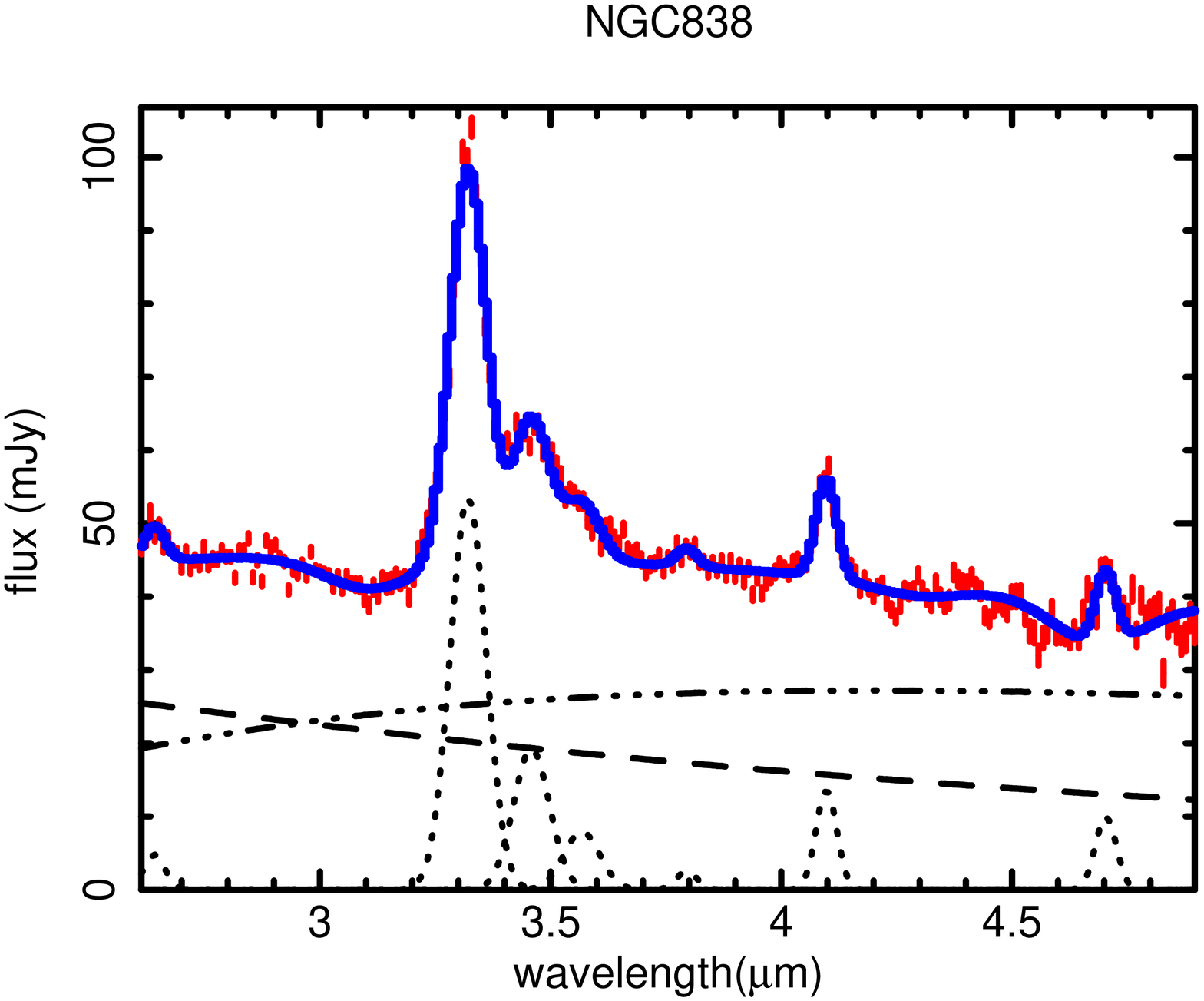}
\includegraphics[angle=0,scale=0.2]{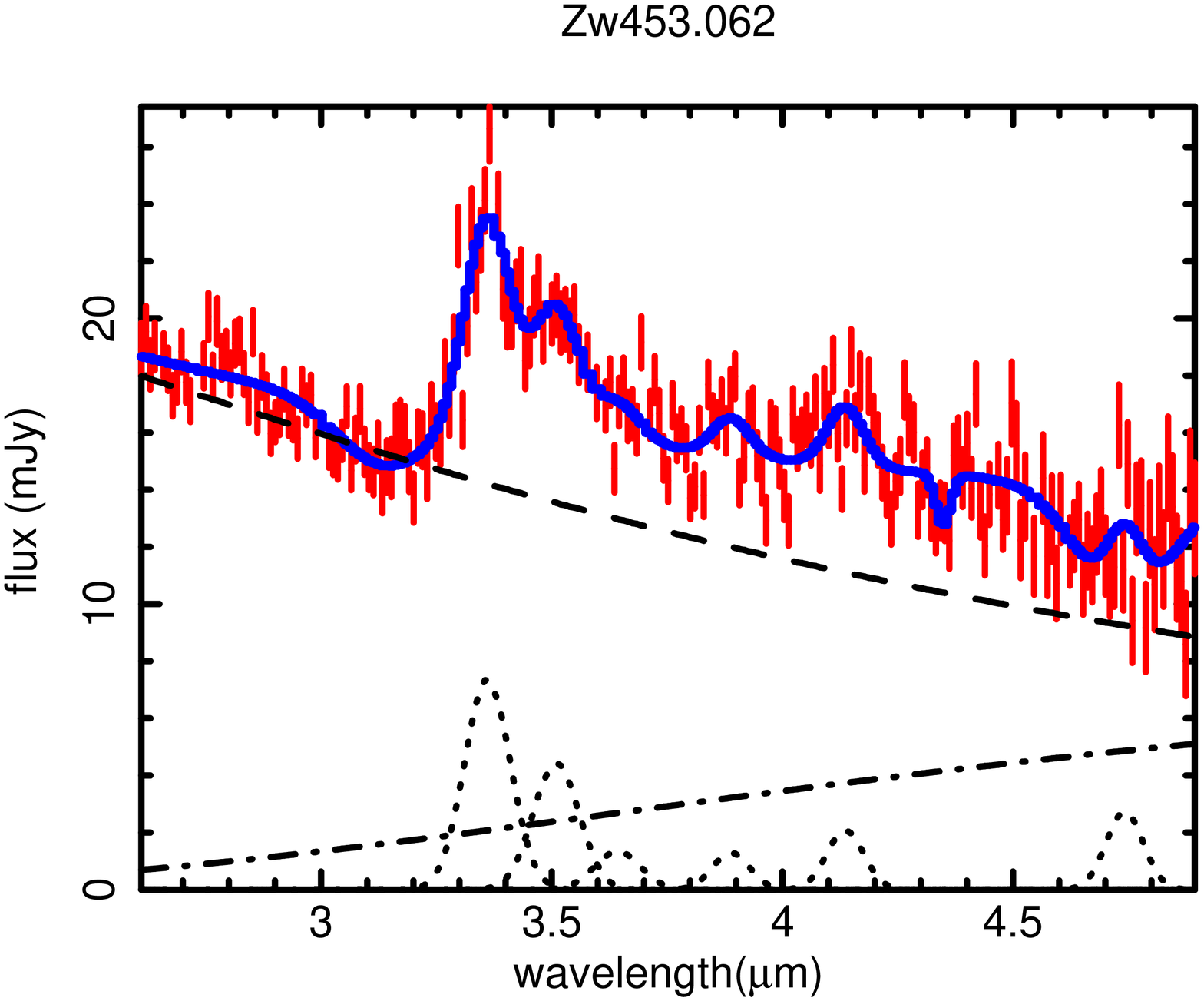}
\end{center}
\caption{(continued)}
\end{figure*}

%

\section{RESULTS and DISCUSSIONS}

\subsection{AKARI spectra and AGN diagnostics}

Figure~2 displays the observed \textit{AKARI} spectra of all the 22
sources (red dots with flux error bars) overplotted with their
best-fit models (blue line). Each line/continuum component is also
plotted in the figure. As noticed, a majority of sources show
3.3~$\mu$m PAH emission features. Table~2 and ~3 summarize the fitting 
results.

For finding buried AGNs, we apply two AGN diagnostics: (1) PAH
emission line and (2) torus-dust continuum. The PAH diagnostic
is based on EW$_{\rm 3.3PAH}$ as discussed in Section~4; we identify a
buried AGN if it shows a small 3.3~$\mu$m PAH equivalent width with
EW$_{\rm 3.3PAH} < 40$~nm. The torus-dust diagnostic is
based on whether or not there is a contribution from a hot dust
component from the torus; we set the criterion of identifying buried
AGN as $T^{(\rm dust)} > 200$~K.

In the left panel of Figure~3, we present the two AGN diagnostic
diagram applied for our sample, where ULIRGs/LIRGs/IRGs are plotted as
pink circles/red triangles/brown squares, respectively. There is a
clear boundary of distribution of the dust temperature around $T^{(\rm
dust)} \sim 200$~K. This supports our criterion that AGNs
should have high dust temperatures ($T^{(\rm dust)} > 200$~K) and 
are well distinguishable from SBs.
Interestingly, many infrared galaxies with large PAH equivalent widths
(EW$_{\rm 3.3PAH} > 40$~nm) show high dust temperatures ($T^{\rm
(dust)} > 200$~K). One reason could be that because of 
the large aperture of \textit{AKARI}/IRC and less luminous nucleus
emission in these galaxies, the spectra contain relatively large
contribution of 3.3~$\mu$m PAH emission in the host galaxies where PAH
molecules are not destroyed by AGN X-ray photons.
This effect works to increase the PAH equivalent width, and hence
the PAH diagnostic could miss buried AGNs. By contrast, the hot-dust
diagnostic is not subject to the aperture effect because the 
continuum emission can be properly decomposed by the spectral fit.
In fact, the torus-dust AGN diagnostic can recover all the buried AGNs
identified by the PAH diagnostic; in other words, there is no source
with a low dust temperature ($T^{\rm (dust)}<200$~K) {\it and} a small
PAH equivalent width (EW$_{\rm 3.3PAH} < 40$~nm) in our sample. This
supports the superiority of using the hot-dust diagnostic for finding
AGNs completely.

\begin{figure*}[t]
\begin{center}
\includegraphics[angle=0,scale=0.65]{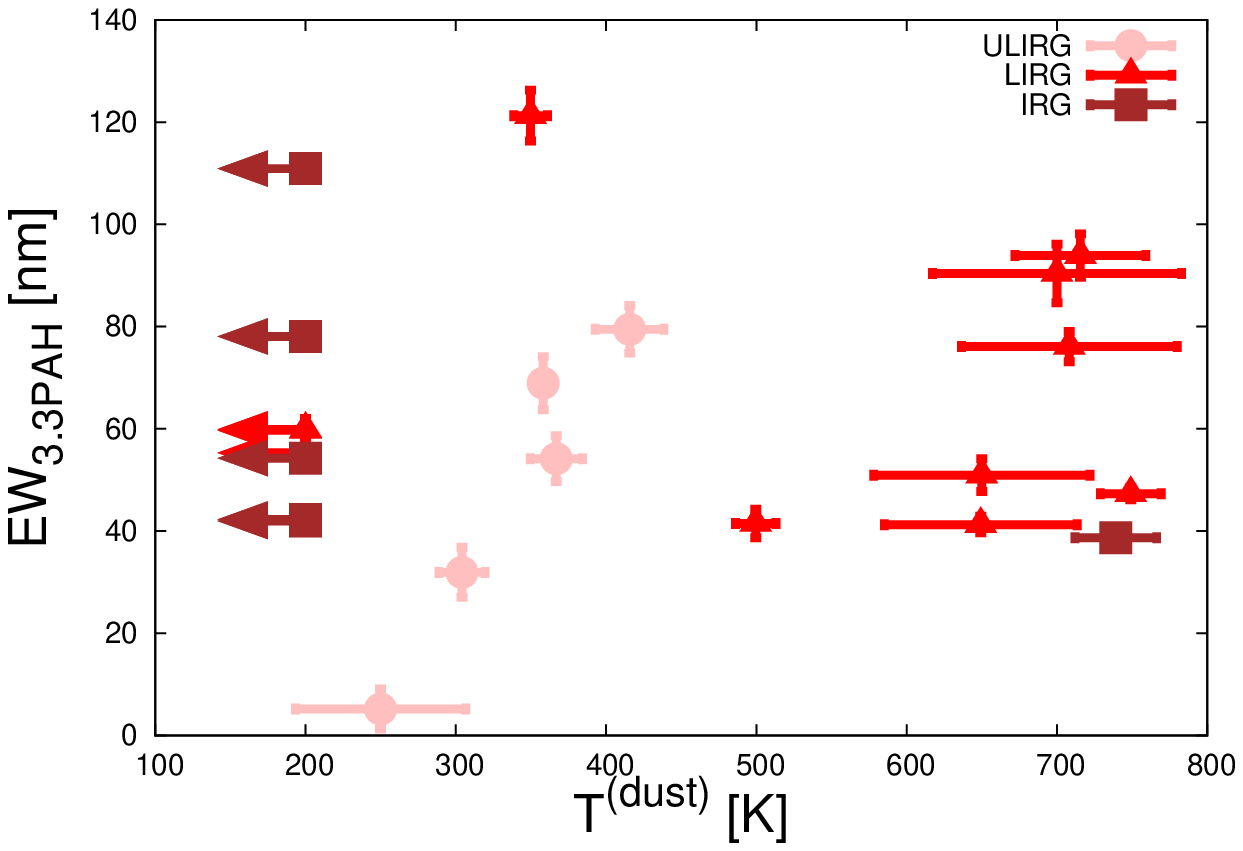}
\includegraphics[angle=0,scale=0.65]{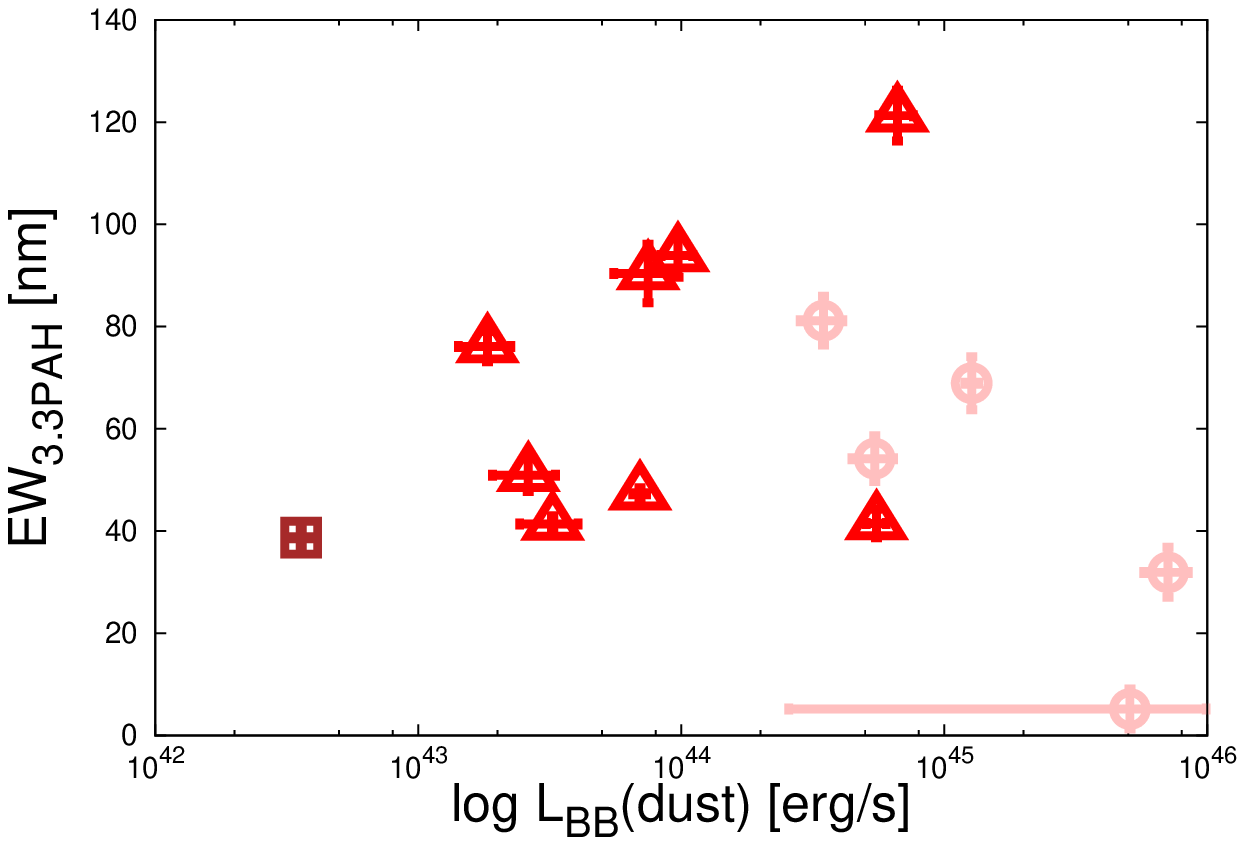}
\\
\caption{
Plot of equivalent width of 3.3~$\mu$m PAH emission
(EW$_{\rm 3.3PAH}$) versus dust temperature $T^{\rm (dust)}$ (left panel)
and infrared AGN luminosity (right panel). Pink 
circles, red triangles, and brown squares represent ULIRGs,
LIRGs, and IRGs, respectively. In the right panel, 
we plot only the sources with $T^{\rm (dust)} > 200$~K
because those with $T^{\rm (dust)} < 200$~K has a great temperature
uncertainty, and therefore the luminosity is given only the small upper
 limit with $10^{40}$~erg s$^{-1}$ (see the discussion in Section~5.1).
\label{fig-3}}
\end{center}
\end{figure*}

The right panel of Figure~3 shows the plot of EW$_{\rm 3.3PAH}$
versus dust luminosity ($L_{\rm BB}^{\rm
(dust)}$) tabulated in Table~2 and Table~3. In
the figure, we plot only the sources with $T>200$~K.
This is because 
the 2.5--5.0~$\mu$m spectral range is not sensitive 
to the dust emission with low temperature $T<100$~K.
In this case, only an upper limit of the dust luminosity can be derived.
This is the reason why the dust luminosity with $T^{(\rm dust)}<200$~K
is very small, 
$L_{\rm BB}^{\rm (dust)}\ll10^{40}$~erg s$^{-1}$.
The figure shows that all the sources with $T^{(\rm dust)}>200$~K have
significant dust emission with $L_{\rm BB}^{\rm (dust)} >
10^{42}$~erg s$^{-1}$.
Thus, the alternative criterion for detecting AGNs can be $L_{\rm BB}^{\rm
(dust)} > 10^{42}$~erg s$^{-1}$ instead of $T^{(\rm dust)}>200$~K . In summary,
we conclude that the hot-dust AGN diagnostic is a better method for
finding buried AGNs in the {\it AKARI} IRC band.

\begin{figure*}[tbp]
\begin{center}
\includegraphics[angle=0,scale=0.65]{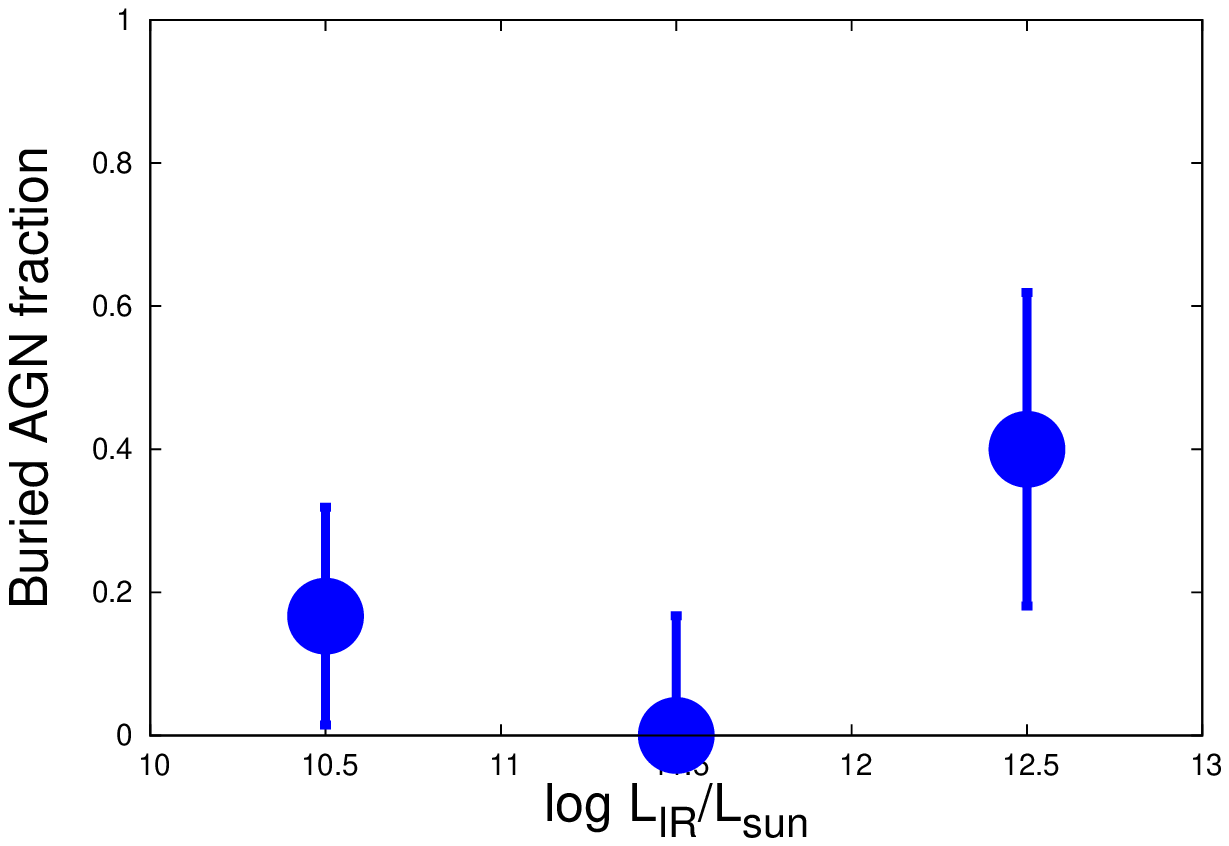}
\includegraphics[angle=0,scale=0.65]{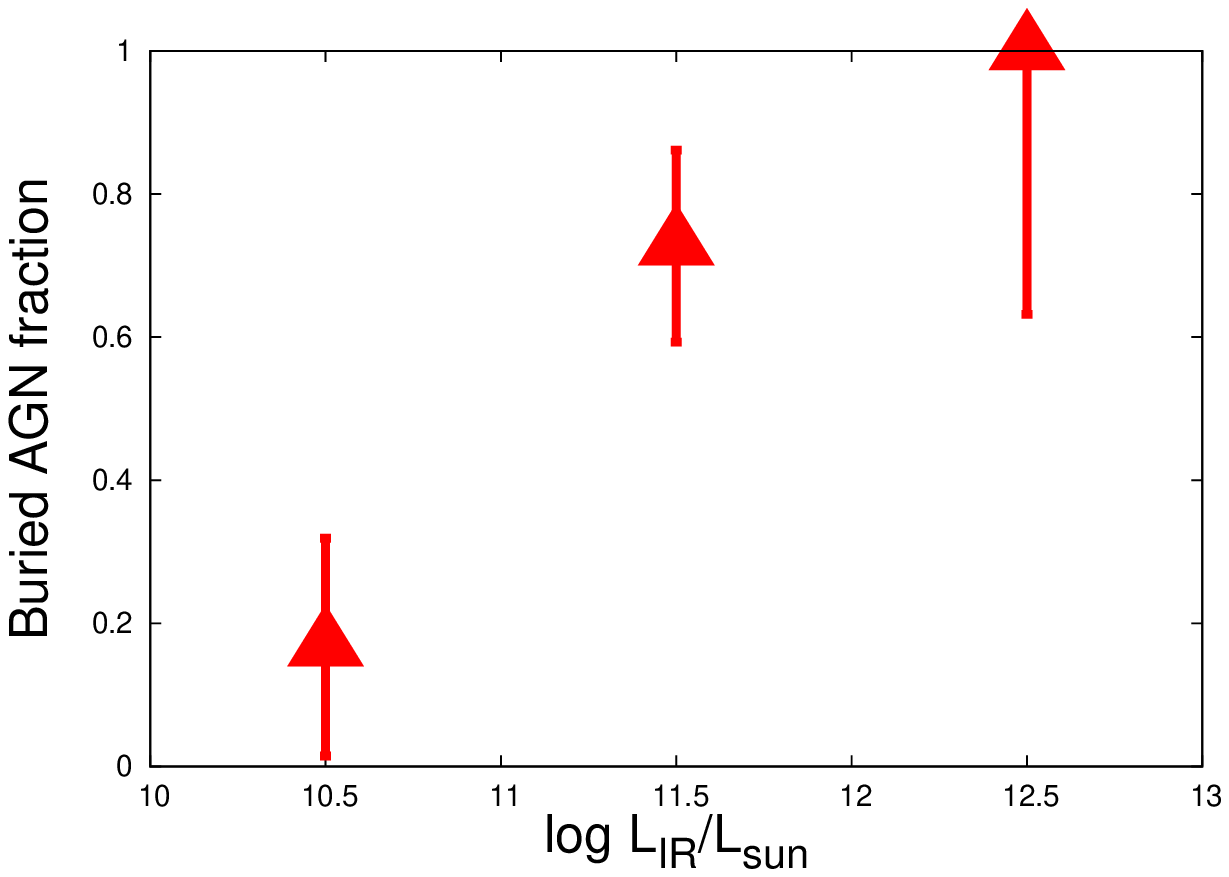}
\\
\caption{
Buried AGN fraction as a function of infrared luminosity. The left 
panel shows the result based on the 3.3~$\mu$m PAH AGN
diagnostic, and the right panel shows that based on the torus-dust AGN
diagnostic.
\label{fig-4}}
\end{center}
\end{figure*}

\subsection{Buried AGN contribution as a function of infrared luminosity}

One motivation of our study is to reveal how frequently buried AGNs
exist in infrared galaxies. Figure~4 shows the fraction of buried AGN
as a function of infrared luminosity. The left and right panels show
the results based on the PAH AGN diagnostic and on the torus-dust AGN
diagnostic, respectively. As discussed above, the latter diagnostic
is more complete than the former for finding buried AGNs from infrared
galaxies. Therefore, we discuss the result of the right panel hereafter.

In total, we detect 14 buried AGNs out of the 22 infrared galaxies
without any AGN signs in the optical band. We also find a clear trend
that the fraction of buried AGN increases with infrared luminosity.
While only 17\% (1 out of 6 sources) of the IRGs contain buried
AGNs, the U/LIRGs contain them almost ubiquitously ($8/11=72$\% for
LIRGs and $5/5=100$\% for ULIRGs). Our results for U/LIRGs are
consistent with those obtained by \cite{ima10} within the statistical
uncertainties. Although most of the IRGs are 
selected from the 12~$\mu$m galaxy catalog,
which is sensitive to torus-dust emission,
an AGN is detected only from one source out of the five targets.

The high AGN fraction in U/LIRGs may be related with their high merger
rates. There is strong observational evidence that most of U/LIRGs
have experienced merging or strong galaxy-galaxy interactions
\citep{vei02}. The mergers can induce strong SB within the galaxies, which
also make galactic gas fall into the galactic center and trigger AGN
activity obscured by dust \citep{san88,mih96,dim05,hop06}. \cite{san04} show
that infrared galaxies with signs of recent merger drastically
increase at $L_{\rm IR} \sim 10^{11.2} L_{\odot}$ toward a higher
luminosity range. This result is in accordance with our finding that
the buried AGN fraction drastically increases from $L_{\rm IR} \sim
10^{11} L_{\odot}$.

\begin{figure}[tbp]
\begin{center}
\includegraphics[angle=0,scale=0.7]{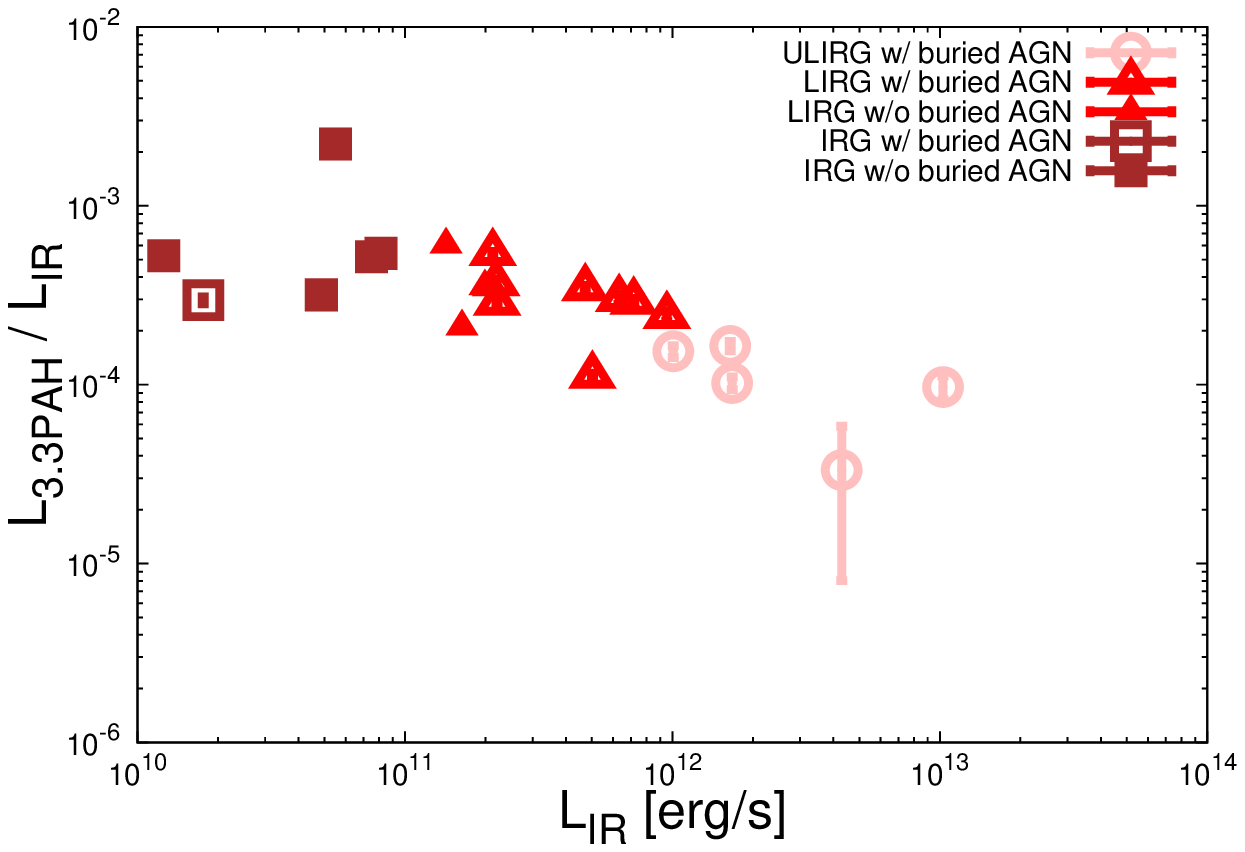}
\\
\caption{
Ratio of $L_{\rm 3.3PAH}/L_{\rm IR}$ as a function of infrared
luminosity. Open circles/triangles/squares represent ULIRGs/LIRGs/IRGs
with AGN signs, respectively. Filled triangle/square represents
LIRGs/IRGs without AGN signs, respectively. 
Note that all ULIRGs in our sample are diagnosed to have buried AGNs.
\label{fig-5}}
\end{center}
\end{figure}

We confirm the dependence of the AGN fraction on luminosity by
investigating the ratio between the 3.3~$\mu$m PAH luminosity ($L_{\rm
3.3PAH}$) and total infrared luminosity $L_{\rm IR}$. Pure SB galaxies
generally have a constant ratio $L_{\rm 3.3PAH} / L_{\rm IR} \sim
10^{-3}$. If an AGN exists inside the galaxy, the ratio becomes
smaller because X-rays from the AGN destroy PAH molecules and the
infrared continuum luminosity is increased by the torus-dust
emission. Figure~5 plots $L_{\rm 3.3PAH} / L_{\rm IR}$ as a function
of infrared luminosity for our sample. As noted, it decreases with
infrared luminosity, especially at $L_{\rm IR} > 2 \times 10^{11}
L_{\odot}$. We also find that the sources with small $L_{\rm 3.3PAH} /
L_{\rm IR}$ ratios match to those with buried AGN signs, supporting
the above idea. Note that there is a possibility that this trend may
be caused by the destruction of small PAH molecules by major mergers
\citep{yam13} rather than by AGNs. Nevertheless, since all the sources with
$L_{\rm 3.3PAH} / L_{\rm IR} < \sim10^{-4}$ show buried AGN signs, the
conservative criterion $L_{\rm 3.3PAH} / L_{\rm IR} < 10^{-4} $ could
be phenomenologically used as another diagnostic for finding buried
AGNs.

\subsection{Luminosity contribution of buried AGN to the total infrared luminosity}

Our second motivation is to determine the energetic importance of AGNs
in these infrared galaxies. 
As discussed in Section~5.1, the luminosity contribution 
from dust blackbody component ($L_{\rm BB}^{(\rm dust)}$) predominantly originates from AGN-heated
dust, and the luminosity contribution of SB-heated dust to $L_{\rm BB}^{(\rm dust)}$ is negligible. 
Thus, using the best-fit blackbody parameters
obtained by the spectral fit, we can estimate the total luminosity
from AGN-heated dust and hence its ratio to the total infrared
luminosity. 
Figure~6 shows the averaged fraction of AGN power
to the total infrared luminosity for IRGs, LIRGs, and ULIRGs. As
noticed, the AGN contribution increases with infrared
luminosity, while the values are quite small,  $\sim 0.9 \pm 0.8$\% in IRGs, $\sim 7.4 \pm 3.3$\%
in LIRGs, and $\sim 19.1 \pm 5.0$\% in ULIRGs. This suggests that the
bulk of the infrared emission originates from SB, not from AGNs in
these galaxies. 
One caveat for this result is that we cannot completely 
exclude the possibility that the estimated luminosity of $L_{\rm
BB}^{(\textsc{Hii})}$ might be partially originated from
the torus-dust if it has an extreme temperature distribution.
We thus 
calculate the averaged $L_{\rm BB}^{(\textsc{Hii})}/L_{\rm
IR}$ ratio to check if it could have a significant contribution.
We find $L_{\rm BB}^{(\textsc{Hii})}/L_{\rm IR} \sim 7$\%, 
suggesting that the main results above would not be affected, even if
the observed very hot ($>1000$~K) dust component
were totally due to the AGN emission.

Another caveat for interpreting this result is the
selection bias for our sample, which consists of optically non-Seyfert
galaxies. Previous studies based on samples including optically
Seyfert AGNs found larger average AGN contribution to the total
infrared luminosity, 10--15\% in LIRGs \citep{pet11,alo12} and
15--40\% in ULIRGs \citep{vei09, nar10, ris10}. 

Our results are in good agreement with those by \cite{lee12} within
the errors, who obtained the AGN to total luminosity ratio of 6--8\%
in LIRGs and 11--19\% in ULIRGs, using a similar sample to ours that
preferentially includes optically non-Seyfert infrared galaxies. They
calculated the AGN contribution by performing SED fit with the
\textsc{Decompir} package \citep{mul11}. This method requires infrared
SED data covering the far-infrared band, while our method only uses
the \textit{AKARI} 2.5--5~$\mu$m spectra. This supports that our
simple assumption in the spectral model (ie, single blackbody for dust
emission) works well in estimating the AGN luminosity.
This simple method will have a great advantage in the era of \textit{JWST},
because we can apply our method to high-$z$ galaxies by observing the
rest 2.5--5.0~$\mu$m band (5.0--10.0~$\mu$m at $z=1$ 
and 7.5--15.0~$\mu$m at $z=2$) with \textit{JWST}/MIRI (5--28~$\mu$m).
Other methods based on the \textit{Spitzer} bandpass (5--35~$\mu$m) 
would have a difficulty in studying high-$z$ objects; for instance,
the observed spectral range corresponds to 15--105~$\mu$m at $z=2$, 
which only the \textit{SPICA} mission can cover.
Thus, our method is achievable for studying infrared galaxies in 
the distant universe even before the launch of \textit{SPICA} (from 2025--).

\begin{figure}[tbp]
\begin{center}
\includegraphics[angle=0,scale=0.7]{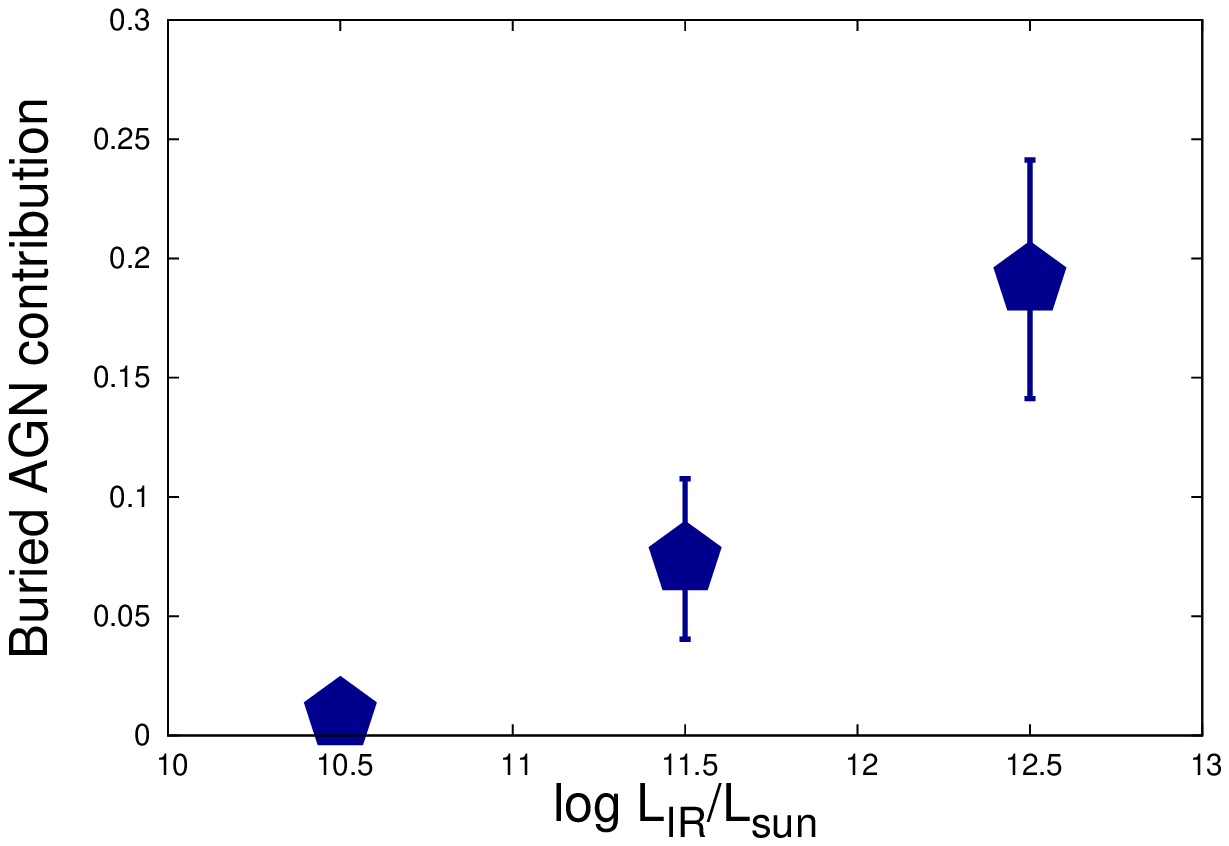}
\\
\caption{
Energy contribution of buried AGNs to the total infrared luminosity 
as a function of infrared luminosity. 
\label{fig-6}}
\end{center}
\end{figure}

Combining our result with the total infrared luminosity density
produced by U/LIRGs in the local universe ($\Omega_{\rm IR}^{(\rm
LIRG)} \sim 5.9 \times 10^{6}$~$L_{\odot}$/Mpc$^3$ and $\Omega_{\rm
IR}^{(\rm ULIRG)} \sim 1.4 \times 10^{5}$~$L_{\odot}$/Mpc$^3$)
\citep{got11}, we estimate the local AGN luminosity density in the
infrared band ($\Omega_{\rm IR}^{\rm (AGN)}$).
We find $\Omega_{\rm IR}^{\rm (AGN)} = 4.4 \times
10^5 L_{\odot}$/Mpc$^{3}$ for LIRGs and $\Omega_{\rm IR}^{\rm (AGN)} =
2.6 \times 10^4 L_{\odot}$/Mpc$^{3}$ for ULIRGs. 
These values are about an order of magnitude lower than the estimate by
\cite{got11}. This is because
\cite{got11} simply assumed that the infrared luminosity of galaxies
classified as AGNs entirely originates from AGNs, while we
quantitatively take into account the fraction of AGN contribution to the total
infrared luminosity of infrared galaxies.
Figure~7 shows the infrared luminosity density as a function of
redshift derived by \cite{got10,got11} together with our results of AGN
the infrared luminosity density.
In the era of \textit{JWST}, we can fill in the AGN
infrared luminosity density at high redshifts thanks to their improved
sensitivity in the mid-infrared band.

\begin{figure}[tbp]
\begin{center}
\includegraphics[angle=0,scale=0.7]{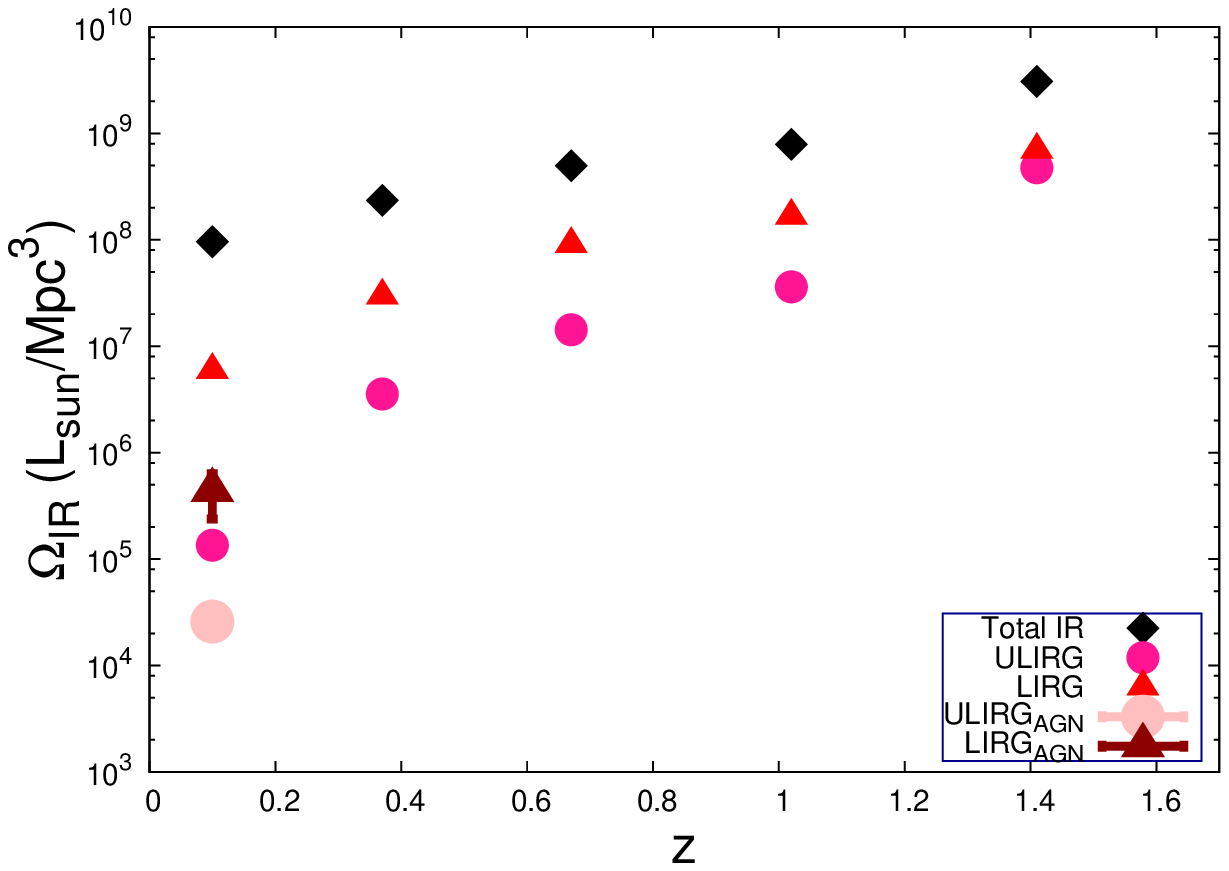}
\\
\caption{
Comoving infrared luminosity density as a function of redshift. Black
filled diamonds represents the infrared luminosity density
($\Omega_{\rm IR}$) of all galaxies, pink filled circles
that of ULIRGs ($\Omega_{\rm IR}^{\rm (ULIRG)}$), and red filled
triangles that of LIRGs ($\Omega_{\rm IR}^{\rm (LIRG)}$),
taken from \cite{got10,got11}. Pale-pink filled circles and brown
filled triangles represent the AGN infrared luminosity density of
ULIRGs and that of LIRGs estimated by our work.
\label{fig-7}}
\end{center}
\end{figure}

\subsection{Evolutionary track of IR galaxies}

Comparison between the starburst and AGN activity gives crucial
information to understand the process of galaxy-SMBH co-evolution.
For optically selected Seyfert galaxies and QSOs, many authors found
that the luminosity of star formation ($L_{\rm SF}$) and that of AGN
($L_{\rm bol}^{\rm (AGN)}$) are well correlated with each other
\citep{net09, oi10,ima11}. 
We can derive a SF luminosity as
$L_{\rm SF} = L_{\rm IR} - L_{\rm BB}^{\rm (dust)}$.
The AGN-heated dust luminosity ($L_{\rm BB}^{(\rm dust)}$) summarized in
Table 3 can be converted to an AGN bolometric luminosity ($L_{\rm
bol}^{(\rm AGN)}$) by using two relations given in 
\cite{gan09} and \cite{mar12}, respectively, 
\begin{align} 
\log L_{\rm 2-10~keV} &= 0.90 \log L_{\rm MIR}^{(\rm AGN)} + 4.09\\ 
\log L_{\rm bol}^{(\rm AGN)} \sim \log L_{\rm disk} &= 1.18 \log L_{\rm
2-10~keV} -6.68.
 \end{align}
Here $L_{\rm disk}$ represents the intrinsic luminosity of the 
accretion disk integrated in the optical to X-ray band. 
Assuming $L_{\rm MIR}^{(\rm AGN)} \sim L_{\rm BB}^{(\rm dust)}$,
we obtain
\begin{align}
\log L_{\rm bol}^{(\rm AGN)} \sim 1.06 \log L_{\rm BB}^{(\rm dust)} - 1.85.
\end{align}
Noted that the estimated bolometric AGN luminosity ($L_{\rm bol}^{(\rm AGN)}$)
derived in Equation~(10) can be overestimated. The relations from Equation~(8) to (10) are derived
from the sample of non-hidden Seyfert AGNs, while our sources are
infrared galaxies with highly obscured (=hidden) AGNs. Therefore, our sources in this study
 could have higher covering fraction of dusty-torus than those of Seyfert AGNs.
 This means the ratio of mid-infrared over bolometric luminosity in our sources
 could be larger than those of Seyfert AGNs.

\begin{figure*}[tbp]
\begin{center}
\includegraphics[angle=0,width=90mm]{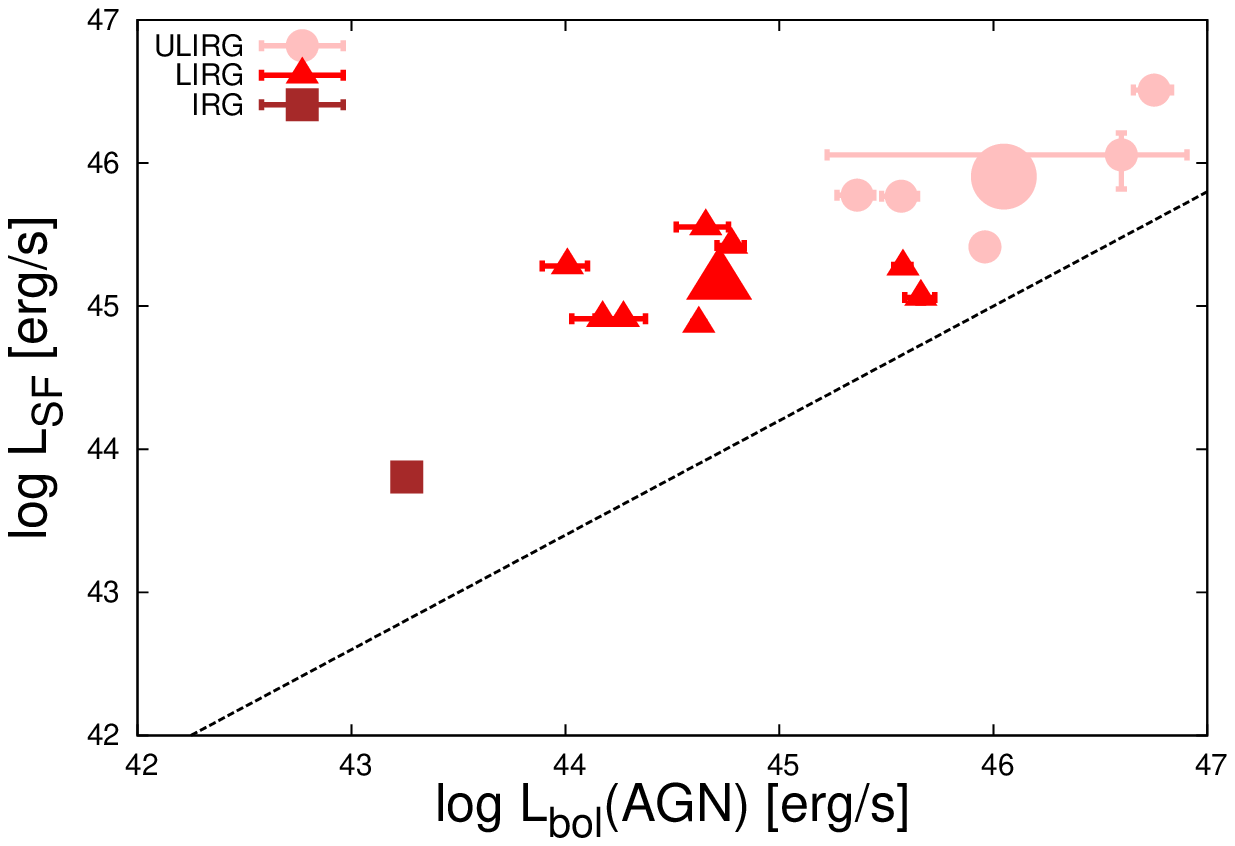}
\includegraphics[angle=0,width=85mm]{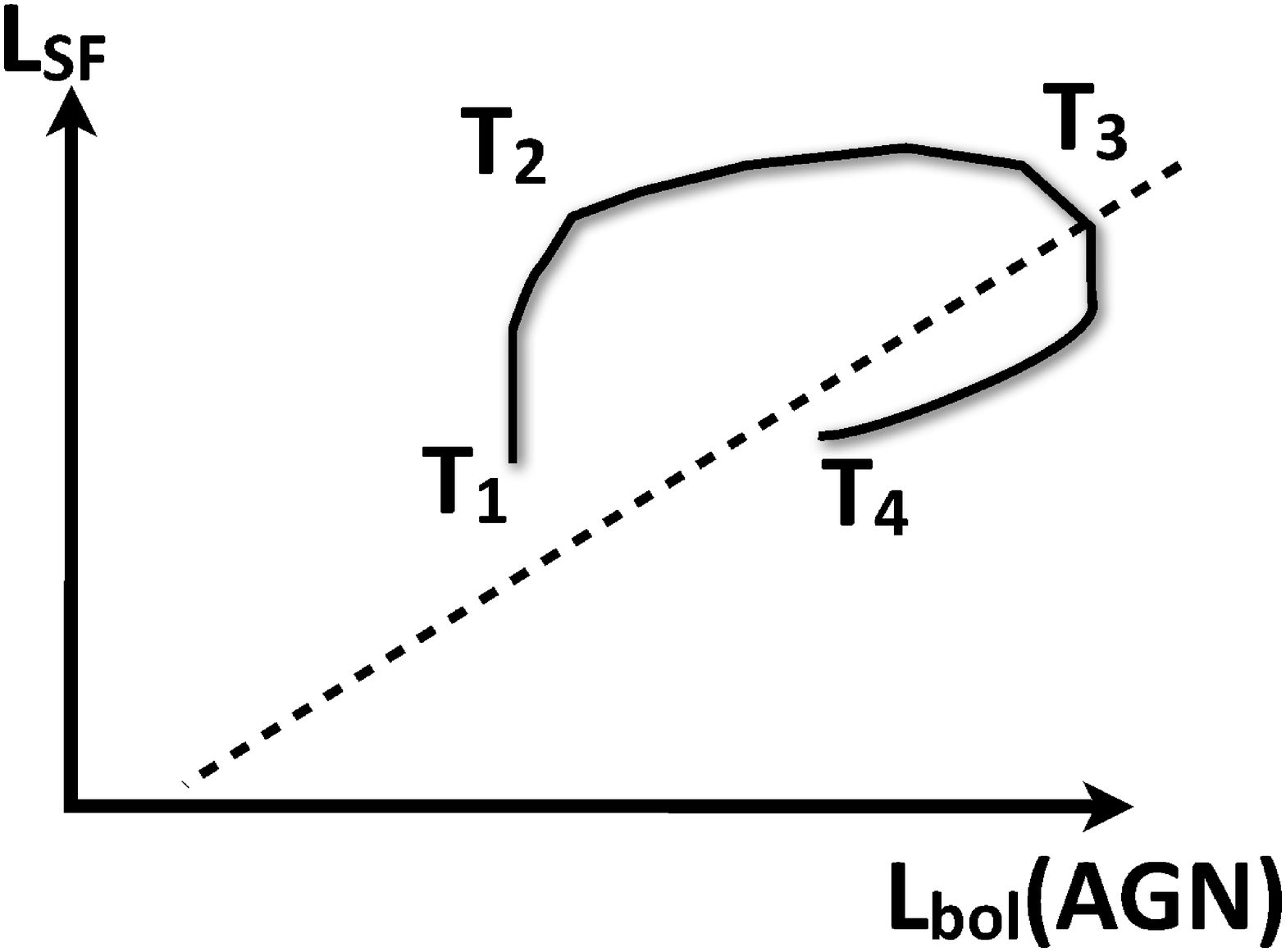}
\caption{
(Left): Plot of SF Luminosity ($L_{\rm SF}$) versus AGN bolometric luminosity
($L_{\rm bol}^{\rm (AGN)}$). The dotted line represents the correlation
line for optically selected Seyfert galaxies taken from
\cite{net09}. All symbols and colors are the same as Figure~3.
(Right) A cartoon illustrating a scenario of AGN evolution from a pure
SB galaxy to an unburied AGN phase. $T_{1}$ represents the evolutional
stage when SB is triggered by merger, $T_2$ when AGN activity is
induced by strong SB, $T_3$ when SB activity starts to decrease due to
the shortage of gas and feedback from the AGN, and $T_4$ when both SB
and AGN activities become weaker. \label{fig-8}}
\end{center}
\end{figure*}

Figure~8 shows the $L_{\rm SF}$ versus $L_{\rm bol}^{(\rm AGN)}$
relations obtained from our sample. The average luminosities of U/LIRG
are also plotted in the same figure. The dotted line corresponds to
the linear correlation obtained from optically selected Seyfert
galaxies obtained by \cite{net09}. Interestingly, all the sources
except two (IRAS F07353+2903 and ESO~286-IG19) are located above the Seyfert line.
This result may reflect the difference in the evolutional stage of
galaxies, from a pure SB phase (U/LIRGs) to an unburied AGN phase
(Seyferts). As discussed in Section 5.2, U/LIRGs have recently
experienced or are facing mergers, which result in strong SB activity
from the epoch $T_1$ to $T_2$ in Figure~8.  The SB also triggers AGN
activity, which continues from $T_2$ to $T_3$. Finally, the SB
activity is gradually weakened due to the shortage of gas and feedback
from the AGN, and the supply of matter from the host galaxy onto the
central engine also decreases, from $T_3$ to $T_4$. In this framework
of galaxy evolution, the evolutional stage of ULIRGs should be a later
phase of mergers than that of LIRGs. The picture is in good agreement
with observations that the morphology of ULIRGs are more compact than
that of LIRGs, suggesting that ULIRGs have just finished the merger
process while that of LIRGs is still on-going.  This scenario suggests
that infrared galaxies containing buried AGNs could be an earlier
evolutional stage of AGN, which will be evolved to normal (unburied)
Seyferts/QSOs. 
This scenario becomes much more reasonable if our bolometric AGN luminosity ($L_{\rm bol}^{(\rm AGN)}$)
is overestimated as discussed in the previous paragraph because all the points are shifted to the left in the figure.
While our sample size is still small, the new AGN
diagnostics developed in this paper can be applied to other
sources. Further studies using a well-defined complete sample of
infrared galaxies should be urged to complete our view of
AGN evolution.

\section{CONCLUSIONS}

We obtained the \textit{AKARI} 2.5--5.0~$\mu$m spectra of 22 infrared
galaxies at $z<0.35$ selected mainly from three infrared catalogs:
bright ULIRG catalog \citep{kla01}, \textit{IRAS} 12~$\mu$m galaxy
catalog \citep{spi02}, and \textit{IRAS} bright galaxy catalog
\citep{soi87, car88}. This band includes unique spectral features for
finding ``buried'' AGNs, the 3.3~$\mu$m PAH emission line and red
continuum spectra originated from AGN torus emission. We perform
detailed spectral fitting to these data by properly modeling the
continuum and emission/absorption features. This enables us to
decompose the continuum spectra into SB and AGN torus-dust
components. The results are summarized as follows.

\begin{enumerate}

\item Using the AGN diagnostics based on the torus-dust continuum
 ($T^{\rm (dust)}>200$~K or $L_{\rm BB}^{\rm (dust)} > 10^{42}$~erg s$^{-1}$)
 and 3.3~$\mu$m PAH emission strength (EW$_{\rm 3.3PAH} < 40$~nm), we
 find that 14 out of the 22 infrared galaxies have buried AGNs. We
 confirm the trend that the buried AGN fraction increases with
 infrared luminosity: $17\pm15$\% for normal IRGs, $72\pm13$\% for
 LIRGs, and $100_{-37}^{+0}$\% for ULIRGs in our sample. This suggests that
 the presence buried AGNs is ubiquitous in U/LIRGs.

 \item The ratio $L_{\rm 3.3PAH} / L_{\rm IR}$ decreases with infrared
 luminosity. This is opposed to the previous reports that $L_{\rm
 3.3PAH} / L_{\rm IR}$ have a constant value \citep{mou90,ima02}. This
 decrease is possibly due to three effects: (1) destruction of PAH
 molecules by X-rays emitted from the AGN, (2) PAH destruction by
 merger processes, and (3) increase of AGN contribution to the total
 infrared luminosity.

 \item The energy contribution from the AGN torus-dust emission to the
 total infrared luminosity also increases with infrared luminosity,
 but it only  reaches $\sim$7\% in LIRGs and $\sim$20\% even in
 ULIRGs. This suggests that majority of the total infrared luminosity
 in U/LIRGs originates from SB, \textit{not} from AGN.

 \item Combining the above results with the luminosity function of
 infrared galaxies derived by \cite{got10,got11}, we estimate the AGN
 luminosity density of $\Omega_{\rm IR}^{\rm (AGN)} = 4.4 \times 10^5
 L_{\odot}$/Mpc$^{3}$ for LIRGs and $\Omega_{\rm IR}^{\rm (AGN)} = 2.6
 \times 10^4 L_{\odot}$/Mpc$^{3}$ for ULIRGs in the local universe.

 \item $L_{\rm SF}$ versus $L_{\rm bol}^{(\rm AGN)}$ luminosity plot
 for our sample shows that infrared galaxies with buried AGNs are
 located above the luminosity correlation line derived from
 optically-selected Seyfert galaxies. We interpret that infrared
 galaxies could be an early phase of the evolutional track of AGNs.

\end{enumerate}

\acknowledgements
We are grateful to Kenichi Yano and Rika Yamada for their help with the
data reduction of \textit{AKARI}. We also thank Claudio Ricci and
Kenta Matsuoka for discussions. The research is based on observations
with \textit{AKARI}, a JAXA project with the participation of
ESA. This work was partly supported by the Grant-in-Aid for JSPS
Fellows for young researchers (K.I.) and for Scientific Research
23540273 (M.I.) and 26400228 (Y.U.).


\begin{thebibliography}{}
\bibitem[Alonso-Herrero et al.(2011)]{alo11} Alonso-Herrero, A., Ramos Almeida, C., Mason, R., et al.\ 2011, \apj, 736, 82

\bibitem[Alonso-Herrero et al.(2012)]{alo12} Alonso-Herrero, A., Pereira-Santaella, M., Rieke, G.~H., \& Rigopoulou, D.\ 2012, \apj, 744, 2

\bibitem[Antonucci(1993)]{ant93} Antonucci, R.\ 1993, \araa, 31, 473

\bibitem[Asensio Ramos \& Ramos Almeida(2009)]{ase09} Asensio Ramos, A., \& Ramos Almeida, C.\ 2009, \apj, 696, 2075

\bibitem[Bauschlicher et al.(2009)]{bau09} Bauschlicher, C.~W., Jr., Peeters, E., \& Allamandola, L.~J.\ 2009, \apj, 697, 311 

\bibitem[Brightman \& Nandra(2011)]{bri11} Brightman, M., \& Nandra, K.\ 2011, \mnras, 413, 1206 

\bibitem[Brightman \& Ueda(2012)]{bri12} Brightman, M., \& Ueda, Y.\ 2012, \mnras, 423, 702

\bibitem[Carico et al.(1988)]{car88} Carico, D.~P., Sanders, D.~B., Soifer, B.~T., et al.\ 1988, \aj, 95, 356 

\bibitem[Di Matteo et al.(2005)]{dim05} Di Matteo, T., Springel, V., \& Hernquist, L.\ 2005, \nat, 433, 604 

\bibitem[Draine(1989)]{dra89} Draine, B.~T.\ 1989, Infrared Spectroscopy in Astronomy, 290, 93

\bibitem[Eguchi et al.(2009)]{egu09} Eguchi, S., Ueda, Y., Terashima, Y., Mushotzky, R., \& Tueller, J.\ 2009, \apj, 696, 1657

\bibitem[Eguchi et al.(2011)]{egu11} Eguchi, S., Ueda, Y., Awaki, H., et al.\ 2011, \apj, 729, 31 

\bibitem[Gandhi et al.(2009)]{gan09} Gandhi, P., Horst, H., Smette, A., et al.\ 2009, \aap, 502, 457

\bibitem[Goto et al.(2010)]{got10} Goto, T., Takagi, T., Matsuhara, H., et al.\ 2010, \aap, 514, A6

\bibitem[Goto et al.(2011)]{got11} Goto, T., Arnouts, S., Inami, H., et al.\ 2011, \mnras, 410, 573 

\bibitem[Goulding et al.(2012)]{gou12} Goulding, A.~D., Alexander, D.~M., Bauer, F.~E., et al.\ 2012, \apj, 755, 5 

\bibitem[Hopkins et al.(2006)]{hop06} Hopkins, P.~F., Hernquist, L., Cox, T.~J., et al.\ 2006, \apjs, 163, 1

\bibitem[Hopkins \& Beacom(2006)]{hop06b} Hopkins, A.~M., \& Beacom, J.~F.\ 2006, \apj, 651, 142 

\bibitem[Hunt et al.(2002)]{hun02} Hunt, L.~K., Giovanardi, C., \& Helou, G.\ 2002, \aap, 394, 873 

\bibitem[Ichikawa et al.(2012)]{ich12} Ichikawa, K., Ueda, Y., Terashima, Y., et al.\ 2012, \apj, 754, 45 

\bibitem[Ichikawa et al.(2012b)]{ich12b} Ichikawa, K., Ueda, Y., Terashima, Y., et al.\ 2012, Torus Workshop, 2012, 109 

\bibitem[Ikeda et al.(2009)]{ike09} Ikeda, S., Awaki, H., \& Terashima, Y.\ 2009, \apj, 692, 608 

\bibitem[Imanishi \& Dudley(2000)]{ima00} Imanishi, M., \& Dudley, C.~C.\ 2000, \apj, 545, 701 

\bibitem[Imanishi(2002)]{ima02} Imanishi, M.\ 2002, \apj, 569, 44 

\bibitem[Imanishi \& Maloney(2003)]{ima03} Imanishi, M., \& Maloney, P.~R.\ 2003, \apj, 588, 165 

\bibitem[Imanishi et al.(2006)]{ima06} Imanishi, M., Dudley, C.~C., \& Maloney, P.~R.\ 2006, \apj, 637, 114

\bibitem[Imanishi et al.(2008)]{ima08} Imanishi, M., Nakagawa, T., Ohyama, Y., et al.\ 2008, \pasj, 60, 489 

\bibitem[Imanishi et al.(2010)]{ima10} Imanishi, M., Nakagawa, T., Shirahata, M., Ohyama, Y., \& Onaka, T.\ 2010, \apj, 721, 1233

\bibitem[Imanishi et al.(2011a)]{ima11} Imanishi, M., Ichikawa, K., Takeuchi, T., et al.\ 2011, \pasj, 63, 447

\bibitem[Imanishi et al.(2011b)]{ima11b} Imanishi, M., Imase, K., Oi, N., \& Ichikawa, K.\ 2011, \aj, 141, 156 

\bibitem[Kishimoto et al.(2011)]{kis11} Kishimoto, M., H{\"o}nig, S.~F., Antonucci, R., et al.\ 2011, \aap, 536, A78

\bibitem[Klaas et al.(2001)]{kla01} Klaas, U., Haas, M., M{\"u}ller, S.~A.~H., et al.\ 2001, \aap, 379, 823

\bibitem[Krolik \& Begelman(1986)]{kro86} Krolik, J.~H., \& Begelman, M.~C.\ 1986, \apjl, 308, L55 

\bibitem[Lee et al.(2012)]{lee12} Lee, J.~C., Hwang, H.~S., Lee, M.~G., Kim, M., \& Lee, J.~H.\ 2012, \apj, 756, 95 

\bibitem[Le Floc'h et al.(2005)]{lef05} Le Floc'h, E., Papovich, C., Dole, H., et al.\ 2005, \apj, 632, 169

\bibitem[Lira et al.(2013)]{lir13} Lira, P., Videla, L., Wu, Y., et al.\ 2013, \apj, 764, 159 

\bibitem[Lu et al.(2003)]{lu03} Lu, N., Helou, G., Werner, M.~W., et al.\ 2003, \apj, 588, 199 

\bibitem[Lutz et al.(1996)]{lut96} Lutz, D., Feuchtgruber, H., Genzel, R., et al.\ 1996, \aap, 315, L269

\bibitem[Magnelli et al.(2011)]{mag11} Magnelli, B., Elbaz, D., Chary, R.~R., et al.\ 2011, \aap, 528, A35

\bibitem[Maiolino et al.(2003)]{mai03} Maiolino, R., Comastri, A., Gilli, R., et al.\ 2003, \mnras, 344, L59 

\bibitem[Marchese et al.(2012)]{mar12} Marchese, E., Della Ceca, R., Caccianiga, A., et al.\ 2012, \aap, 539, A48 

\bibitem[Mihos \& Hernquist(1996)]{mih96} Mihos, J.~C., \& Hernquist, L.\ 1996, \apj, 464, 641 

\bibitem[Moorwood(1986)]{moo86} Moorwood, A.~F.~M.\ 1986, \aap, 166, 4

\bibitem[Mouri et al.(1990)]{mou90} Mouri, H., Kawara, K., Taniguchi, Y., \& Nishida, M.\ 1990, \apjl, 356, L39 

\bibitem[Mullaney et al.(2011)]{mul11} Mullaney, J.~R., Alexander, D.~M., Goulding, A.~D., \& Hickox, R.~C.\ 2011, \mnras, 414, 1082 

\bibitem[Murakami et al.(2007)]{mur07} Murakami, H., Baba, H., Barthel, P., et al.\ 2007, \pasj, 59, 369 

\bibitem[Murphy et al.(2011)]{mur11} Murphy, E.~J., Chary, R.-R., Dickinson, M., et al.\ 2011, \apj, 732, 126

\bibitem[Nardini et al.(2008)]{nar08} Nardini, E., Risaliti, G., Salvati, M., et al.\ 2008, \mnras, 385, L130 

\bibitem[Nardini et al.(2009)]{nar09} Nardini, E., Risaliti, G., Salvati, M., et al.\ 2009, \mnras, 399, 1373 

\bibitem[Nardini et al.(2010)]{nar10} Nardini, E., Risaliti, G., Watabe, Y., Salvati, M., \& Sani, E.\ 2010, \mnras, 405, 2505 

\bibitem[Nenkova et al.(2002)]{nen02} Nenkova, M., Ivezi{\'c}, {\v Z}., \& Elitzur, M.\ 2002, \apjl, 570, L9 

\bibitem[Nenkova et al.(2008a)]{nen08a} Nenkova, M., Sirocky, M.~M., Ivezi{\'c}, {\v Z}., \& Elitzur, M.\ 2008, \apj, 685, 147 

\bibitem[Nenkova et al.(2008b)]{nen08b} Nenkova, M., Sirocky, M.~M., Nikutta, R., Ivezi{\'c}, {\v Z}., \& Elitzur, M.\ 2008, \apj, 685, 160 

\bibitem[Netzer(2009)]{net09} Netzer, H.\ 2009, \mnras, 399, 1907 

\bibitem[Nishiyama et al.(2009)]{nis09} Nishiyama, S., Tamura, M., Hatano, H., et al.\ 2009, \apj, 696, 1407

\bibitem[Ohyama et al.(2007)]{ohy07} Ohyama, Y., Onaka, T., Matsuhara, H., et al.\ 2007, \pasj, 59, 411 

\bibitem[Oi et al.(2010)]{oi10} Oi, N., Imanishi, M., \& Imase, K.\ 2010, \pasj, 62, 1509 

\bibitem[Onaka et al.(2007)]{ona07} Onaka, T., Matsuhara, H., Wada, T., et al.\ 2007, \pasj, 59, 401 

\bibitem[Oyabu et al.(2011)]{oya11} Oyabu, S., Ishihara, D., Malkan, M., et al.\ 2011, \aap, 529, A122 

\bibitem[Petric et al.(2011)]{pet11} Petric, A.~O., Armus, L., Howell, J., et al.\ 2011, \apj, 730, 28 

\bibitem[Ramos Almeida et al.(2009)]{ram09} Ramos Almeida, C., Levenson, N.~A., Rodr{\'{\i}}guez Espinosa, J.~M., et al.\ 2009, \apj, 702, 1127 

\bibitem[Ramos Almeida et al.(2011)]{ram11} Ramos Almeida, C., Levenson, N.~A., Alonso-Herrero, A., et al.\ 2011, \apj, 731, 92 

\bibitem[Risaliti et al.(2006)]{ris06} Risaliti, G., Maiolino, R., Marconi, A., et al.\ 2006, \mnras, 365, 303

\bibitem[Risaliti et al.(2010)]{ris10} Risaliti, G., Imanishi, M., \& Sani, E.\ 2010, \mnras, 401, 197 

\bibitem[Rush et al.(1993)]{rus93} Rush, B., Malkan, M.~A., \& Spinoglio, L.\ 1993, \apjs, 89, 1

\bibitem[Sanders et al.(1988)]{san88} Sanders, D.~B., Soifer, B.~T., Elias, J.~H., et al.\ 1988, \apj, 325, 74 

\bibitem[Sanders \& Mirabel(1996)]{san96} Sanders, D.~B., \& Mirabel, I.~F.\ 1996, \araa, 34, 749

\bibitem[Sanders \& Ishida(2004)]{san04} Sanders, D., \& Ishida, C.\ 2004, The Neutral ISM in Starburst Galaxies, 320, 230 

\bibitem[Sani et al.(2008)]{san08} Sani, E., Risaliti, G., Salvati, M., et al.\ 2008, \apj, 675, 96

\bibitem[Sawicki(2002)]{saw02} Sawicki, M.\ 2002, \aj, 124, 3050 

\bibitem[Siebenmorgen(1993)]{sie93} Siebenmorgen, R.\ 1993, \apj, 408, 218 

\bibitem[Soifer et al.(1987)]{soi87} Soifer, B.~T., Sanders, D.~B., Madore, B.~F., et al.\ 1987, \apj, 320, 238 

\bibitem[Soifer et al.(2001)]{soi01} Soifer, B.~T., Neugebauer, G., Matthews, K., et al.\ 2001, \aj, 122, 1213 

\bibitem[Spinoglio et al.(1995)]{spi95} Spinoglio, L., Malkan, M.~A., Rush, B., Carrasco, L., \& Recillas-Cruz, E.\ 1995, \apj, 453, 616

\bibitem[Spinoglio et al.(2002)]{spi02} Spinoglio, L., Andreani, P., \& Malkan, M.~A.\ 2002, \apj, 572, 105 

\bibitem[Stalevski et al.(2012)]{sta12} Stalevski, M., Fritz, J., Baes, M., Nakos, T., \& Popovi{\'c}, L.~{\v C}.\ 2012, \mnras, 420, 2756

\bibitem[Ueda et al.(2007)]{ued07} Ueda, Y., Eguchi, S., Terashima, Y., et al.\ 2007, \apjl, 664, L79 

\bibitem[Veilleux et al.(1995)]{vei95} Veilleux, S., Kim, D.-C., Sanders, D.~B., Mazzarella, J.~M., \& Soifer, B.~T.\ 1995, \apjs, 98, 171 

\bibitem[Veilleux et al.(1999)]{vei99} Veilleux, S., Kim, D.-C., \& Sanders, D.~B.\ 1999, \apj, 522, 113 

\bibitem[Veilleux et al.(2002)]{vei02} Veilleux, S., Kim, D.-C., \& Sanders, D.~B.\ 2002, \apjs, 143, 315 

\bibitem[Veilleux et al.(2009)]{vei09} Veilleux, S., Rupke, D.~S.~N., Kim, D.-C., et al.\ 2009, \apjs, 182, 628 

\bibitem[V{\'e}ron-Cetty \& V{\'e}ron(2006)]{ver06} V{\'e}ron-Cetty, M.-P., \& V{\'e}ron, P.\ 2006, \aap, 455, 773 

\bibitem[Videla et al.(2013)]{vid13} Videla, L., Lira, P., Andrews, H., et al.\ 2013, \apjs, 204, 23 

\bibitem[Winter et al.(2009)]{win09} Winter, L.~M., Mushotzky, R.~F., Reynolds, C.~S., \& Tueller, J.\ 2009, \apj, 690, 1322

\bibitem[Yamada et al.(2013)]{yam13} Yamada, R., Oyabu, S., Kaneda, H., et al.\ 2013, \pasj~accepted, arXiv:1307.6356

\bibitem[Yuan et al.(2010)]{yua10} Yuan, T.-T., Kewley, L.~J., \& Sanders, D.~B.\ 2010, \apj, 709, 884 

\end{thebibliography}
\end{document}